\documentclass[manuscript]{aastex}
\usepackage{amsmath}

\newcommand{\be}{\begin{equation}}
\newcommand{\ee}{\end{equation}}
\newcommand{\ba}{\begin{eqnarray}}
\newcommand{\ea}{\end{eqnarray}}

\newcommand{\simgt}{\lower 2pt \hbox{$\, \buildrel {\scriptstyle >}\over {\scriptstyle\sim}\,$}}
\newcommand{\simlt}{\lower 2pt \hbox{$\, \buildrel {\scriptstyle <}\over {\scriptstyle\sim}\,$}}
\newcommand{\ls}{\lower 2pt \hbox{$\;\scriptscriptstyle \buildrel<\over\sim\;$}}
\newcommand{\gs}{\lower 2pt \hbox{$\;\scriptscriptstyle \buildrel>\over\sim\;$}}

\slugcomment{To appear in ApJ}

\shorttitle{Soft vs Hard X-ray Microlensing in Chandra Quasars}
\shortauthors{Guerras et al.}

\begin{document}

\title{Extended X-ray Monitoring of Gravitational Lenses with Chandra and Joint Constraints on X-ray Emission Regions}

\author{Eduardo Guerras\altaffilmark{1}, Xinyu Dai\altaffilmark{1}, Shaun Steele\altaffilmark{1}, Ang Liu \altaffilmark{1}, Christopher S. Kochanek\altaffilmark{2}, George Chartas\altaffilmark{3}, Christopher W. Morgan\altaffilmark{4}, Bin Chen\altaffilmark{5}}

\altaffiltext{1}{Homer L. Dodge Department of Physics and Astronomy, The University of Oklahoma,
Norman, OK, 73019, USA}
\email{e.guerras.valera@gmail.com}

\altaffiltext{2}{Department of Astronomy, The Ohio State University, Columbus, OH 43210, USA}
\altaffiltext{3}{Department of Physics and Astronomy, College of Charleston, Charleston, SC 29424, USA}
\altaffiltext{4}{Department of Physics, United States Naval Academy, 572C Holloway Road, Annapolis, MD 21402}
\altaffiltext{5}{Department of Scientific Computing, Florida State University, Tallahassee, FL 32306, USA}

\begin{abstract}

We present an X-ray photometric analysis of six gravitationally lensed quasars, with observation campaigns spanning from $5$ to $14$ years, measuring the total ($0.83-21.8$ keV restframe), soft ($0.83-3.6$ keV), and hard ($3.6-21.8$ keV) band image flux ratios for each epoch. Using the ratios of the model-predicted macro-magnifications as baselines, we build differential microlensing light curves and obtain joint likelihood functions for the average X-ray emission region sizes. Our analysis yields a Probability Distribution Function for the average half-light radius of the X-Ray emission region in the sample that peaks slightly above $1$ gravitational radius and with nearly indistinguishable $68\%$ confidence (one-sided) upper limits of $17.8$ and $18.9$ gravitational radii for the soft and hard X-ray emitting regions, assuming a mean stellar mass of $0.3$ $M_{\odot}$. We see hints of energy dependent microlensing between the soft and hard bands in two of the objects. In a separate analysis on the root-mean-square (RMS) of the microlensing variability, we find significant differences between the soft and hard bands but the sign of the difference is not consistent across the sample. This suggests the existence of some kind of spatial structure to the X-ray emission in an otherwise extremely compact source. We also discover a correlation between the RMS microlensing variability and the average microlensing amplitude.

\end{abstract}
    

    
\section{Introduction}

X-ray emission is one of the defining characteristics of active galactic nuclei (AGN). However, most properties of the X-ray corona are obtained only through spectral analyses, since neither current nor near-future instrumentation can resolve the X-ray emitting regions of AGN. Reverberation mapping and quasar microlensing provide the only probes of the spatial structure of the different AGN components, with the latter better suited to the more compact regions like the X-ray corona or the accretion disk. Reverberation mapping studies have succeeded in mapping the more spatially extended regions such as the broad line regions \citep[e.g., ][]{Bentz2009,Zu2011,Bentz2013,Kollatschny2014}, the dust torus \citep[e.g., ][]{Koshida2014}, and with limited results for accretion disks \citep[e.g., ][]{Shappee2014, Fausnaugh2016}.

Microlensing refers to the micro-arcsecond effects produced by light ray deflections of emision from a background source by foreground stars. It has the advantage over reverberation mapping in that the signal only becomes stronger as the source becomes more compact. The Einstein radius gives a typical scale of
\begin{equation} \label{eq:AE}
 R_{E}=D_{ol}\sqrt{\frac{4GM}{c^{2}}\frac{D_{ls}}{D_{ol}D_{os}}},
\end{equation}
where $M$ denotes the deflector mass and $D_{ol}$, $D_{os}$, and $D_{ls}$ are the angular diameter distances between the observer, lens, and source respectively. If the apparent size of the source is comparable or smaller in size than the Einstein radius, typically a few light-days, the observed flux varies because the magnification changes as the source, lens, and observer move relative to each other \citep[see, e.g., the review by][]{Wambsganss2006}. This makes extragalactic microlensing a unique tool for probing the spatial structure of the central region of quasars, because most AGN components are comparable in size to the Einstein radius or smaller. As a result, microlensing has been successfuly used to obtain size estimates for the broad line region spanning several tens of light days \citep[e.g., ][]{Sluse2012,Guerras2013a}, the accretion disk spanning $\sim$10 light-days \citep[e.g., ][]{Morgan2008,Mediavilla2011a,Jorge2014}, and the X-ray corona of $\sim$1 light-day \citep[e.g., ][]{Dai2010,Pooley2012,Morgan2012,Mosquera2013,Blackburne2014,Blackburne2015,Macleod2015}.

Here, we present updated X-ray light curves for a sample of $6$ lensed quasars with redshifts between $z_{s}=1.3$ and $z_{s}=2.3$. We derive total, soft, and hard energy band light curves in Section~\ref{sec:photometries} and examine them for evidence of microlensing. We perform a simple analysis of several aspects of the microlensing variability in Section~\ref{sec:microlensing_analysis}. In Section~\ref{sec:source_size_estimates} we derive a Probability Distribution Function (PDF) for the average size of the X-ray emitting region in the sample. Section~\ref{sec:discussion} presents a summary of the results.
We assume a flat $\Lambda$CDM cosmology with $H_0 = 70~{\rm km~s^{-1} Mpc}^{-1}$, $\Omega_m = 0.3$, and $\Omega_\Lambda=0.7$.


\section{Image Models and Photometry}\label{sec:photometries}

\begin{figure}
\epsscale{1.00}
\plotone{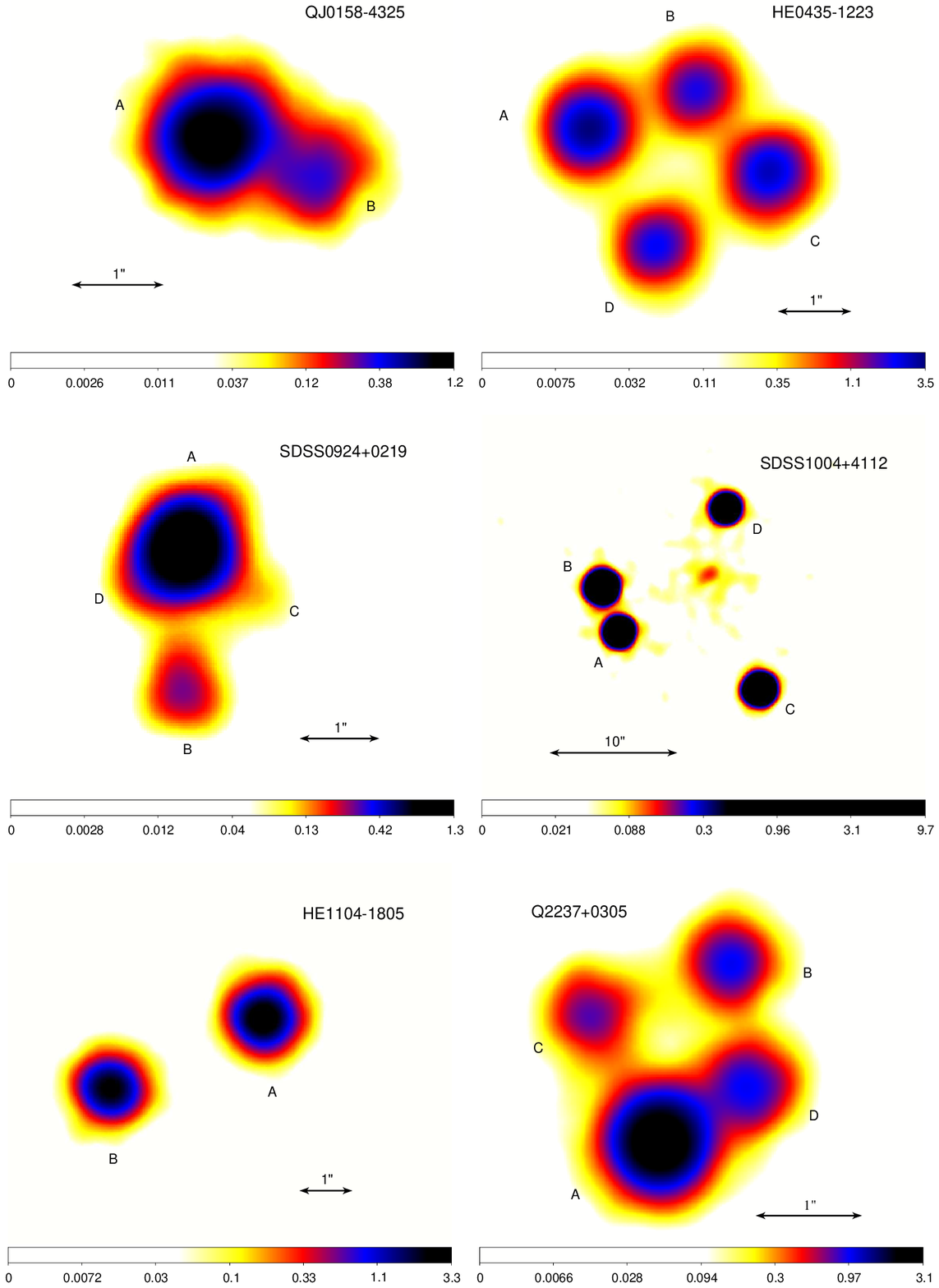}
\caption{Stacked images of the objects in the sample. Note that the angular scale is not uniform.\label{fig:stacked_all}}
\end{figure}

We observed five gravitationally lensed quasars with {\emph{Chandra}}/ACIS \citep{Weisskopf2002,Garmire2003} in Cycles 14--16 for a total exposure time of 810~ks. For each object, we obtained 6--8 sparse monitoring observations over a period of 2--3 years. We also include any archival data for the five systems and an aditional one in our analysis, adding up to six gravitational lenses. All data (both new and archival) were re-calibrated and processed with the latest CIAO 4.7 software\footnote{http://cxc.harvard.edu/ciao/}. Figure \ref{fig:stacked_all} presents stacked images of the six systems, and Table \ref{tab:summaryofdata} summarizes their basic properties. Only in SDSS 1004+4112 are the lensed images easily resolved by {\emph{Chandra}} so that we could do a simple aperture photometry. For this system, we also correct for the background emission from the lensing cluster using arc-shaped regions opposed to each image with respect to the center of the X-ray cluster. For the rest of the systems, we used PSF fitting based on the known relative positions of the images as they are listed in the CASTLES website\footnote{https://www.cfa.harvard.edu/castles/} (see references therein) to model the image fluxes, because the typical angular image separation is not much bigger than the $\sim0\arcsec.5$ arcsec on-axis PSF of {\emph{Chandra}} and aperture photometry would be contaminated by the flux from nearby images. The details of our approach to PSF fitting and photometry can be found in \cite{Chen2012} who analysed data from our previous observational campaigns.

Unlike \cite{Chen2012}, we used a fixed rest frame energy boundary of $3.6$ keV to define the rest frame soft ($0.8-3.6$ keV) and hard ($3.6-21.8$ keV) bands. This both produces comparable count rates for each band and leads to a well-defined combined analysis in Section \ref{sec:source_size_estimates}. Count rates are background subtracted and corrected for both Galactic absorption and absorption by the lens galaxy. To estimate the latter, we closely followed the steps detailed by \cite{Chen2012}. We fit a simple power law with a Gaussian emission line model to the stacked spectra of individual images. The absorption of the lens galaxy was allowed to vary independently in the fit for each image, while the power law index was assumed to be the same for all the images of each quasar. Further details of the absorption correction in our data will be presented in a companion paper (Steele et al.\ 2017, in preparation) which focuses on the spectral analysis of the sample. Tables \ref{tab:0158count} to \ref{tab:2237count} present the absorption-corrected count rates for each lens.

We also set limits on the flux of any central image found by combining all epochs. This is not feasible for SDSS1004+4112, which is known to have a central image \citep[see][]{Inada2008}, because this is also where the cluster X-ray emission peaks \citep[see][]{Ota2006}. Relative to the mean flux of the faintest observed image, the upper limits on the relative flux of any central image are $0.64, 0.009, 1.0, 0.036, 0.019$ for QJ 0158$-$4325, HE 0435$-$1223, SDSS 0924+0219, HE 1104$-$1805, and Q 2237+0305, respectively, at 68\% confidence level. Based on the expected flux ratios of central images \citep[see][]{Keeton2003}, only the limit for HE 0435$-$1223 is strong enough to be useful as an upper limit on the central surface mass density of the lens model.


\section{Microlensing Analysis}\label{sec:microlensing_analysis}
\begin{figure}
\epsscale{.80}
\plotone{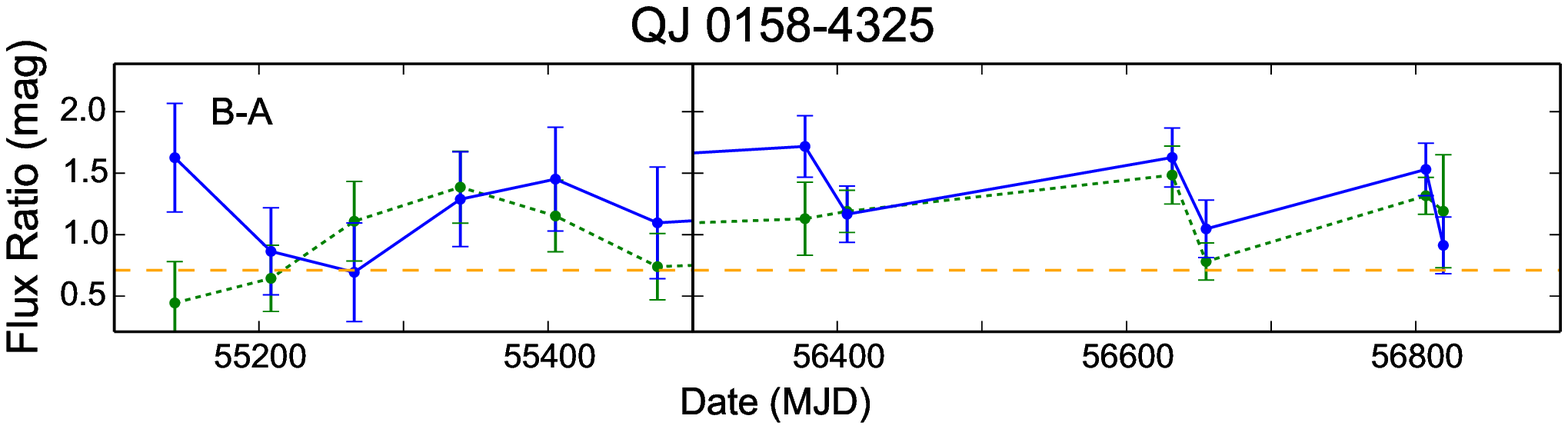}
\caption{Flux ratios for QJ 0158$-$4325 on a magnitude scale. Continuous blue (dashed green) curves show the hard (soft) emission. The orange dashed horizontal lines represent the baseline ratios.\label{fig:0158}}
\end{figure}

\begin{figure}
\epsscale{.80}
\plotone{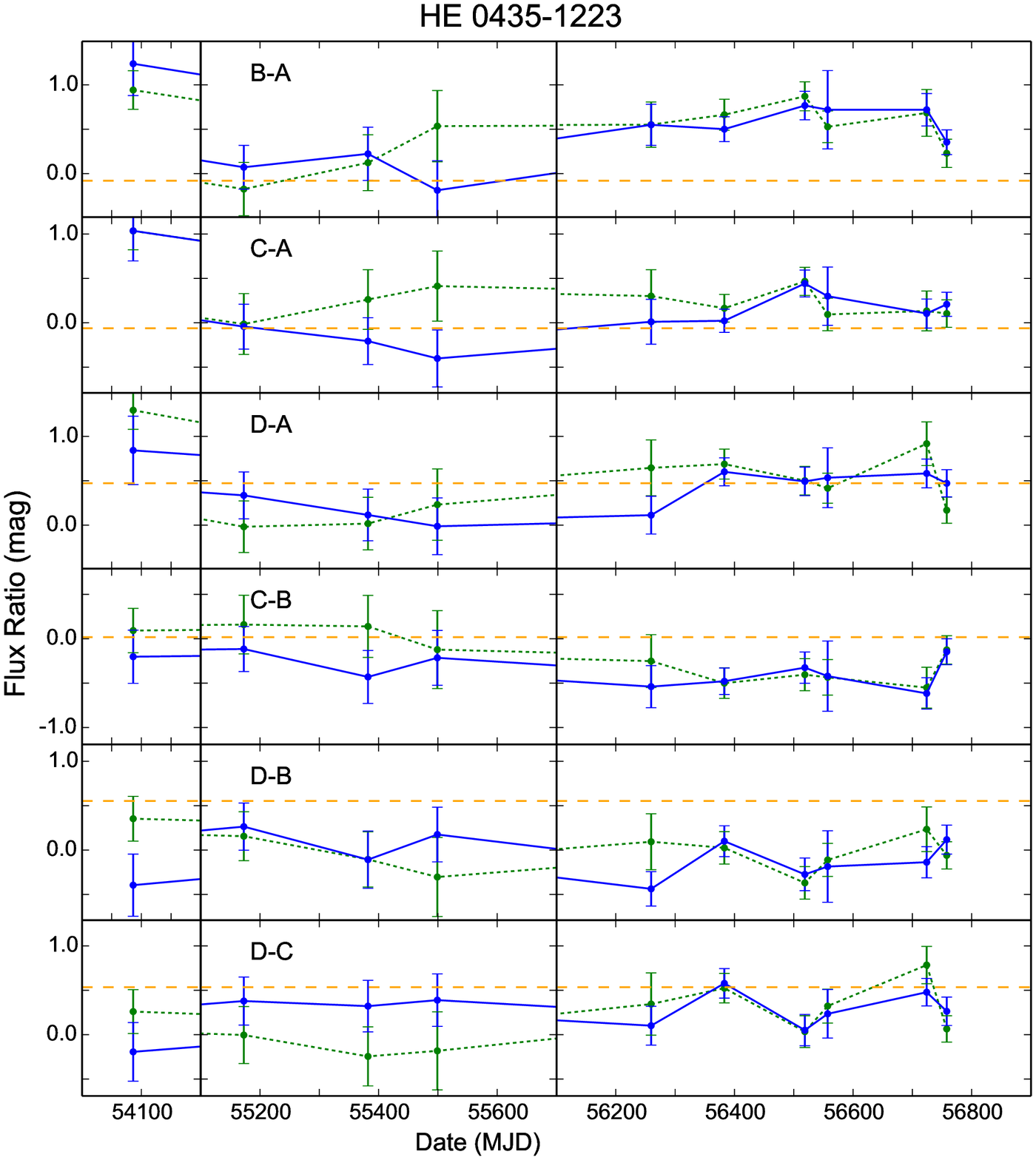}
\caption{Flux ratios for HE 0435$-$1223 on a magnitude scale. Continuous blue (dashed green) curves show the hard (soft) emission. The orange dashed horizontal lines represent the baseline ratios.\label{fig:0435}}
\end{figure}

\begin{figure}
\epsscale{.80}
\plotone{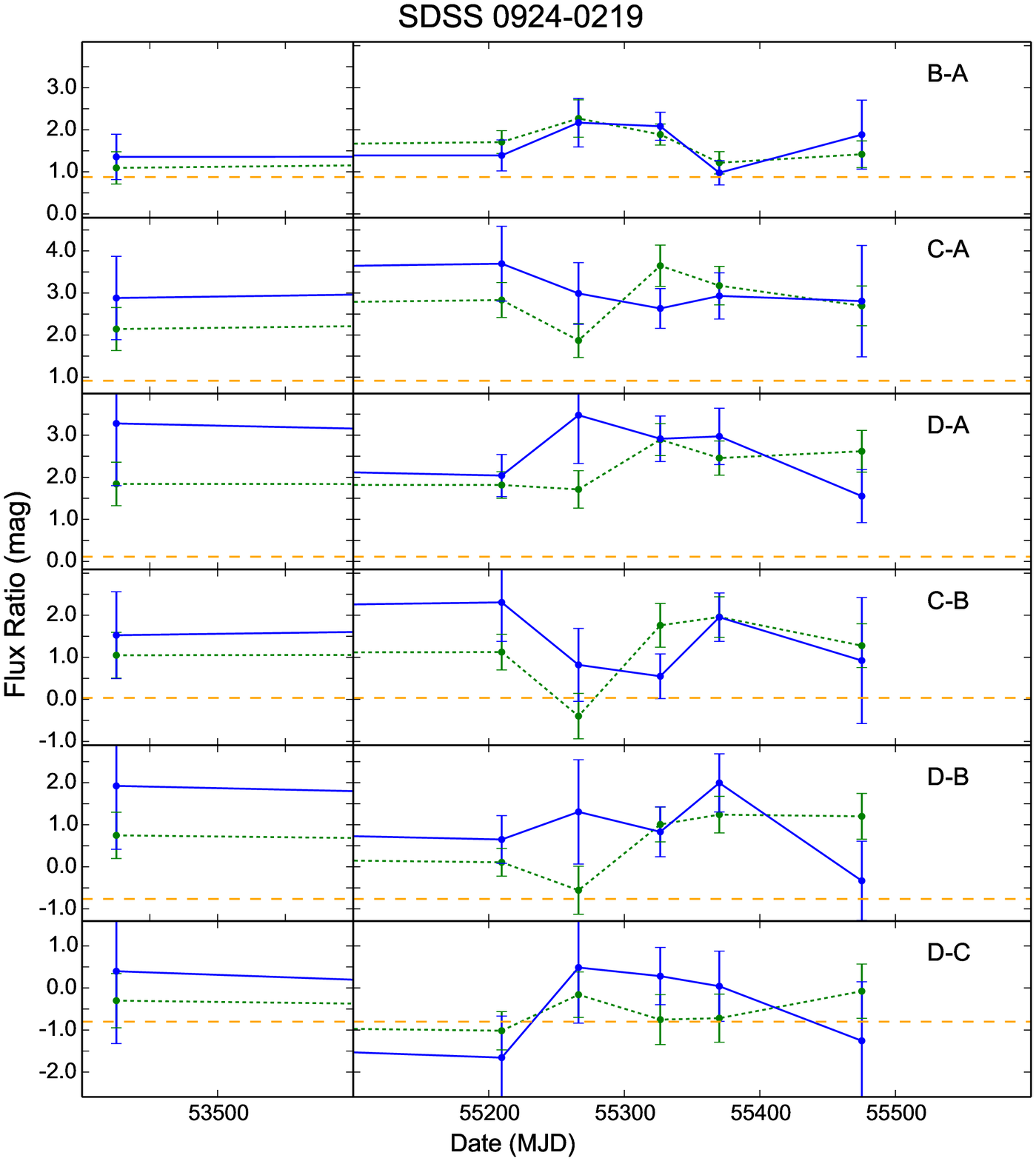}
\caption{Flux ratios for SDSS 0924+0219 on a magnitude scale. Continuous blue (dashed green) curves show the hard (soft) emission. The orange dashed horizontal lines represent the baseline ratios.\label{fig:0924}}
\end{figure}

\begin{figure}
\epsscale{.80}
\plotone{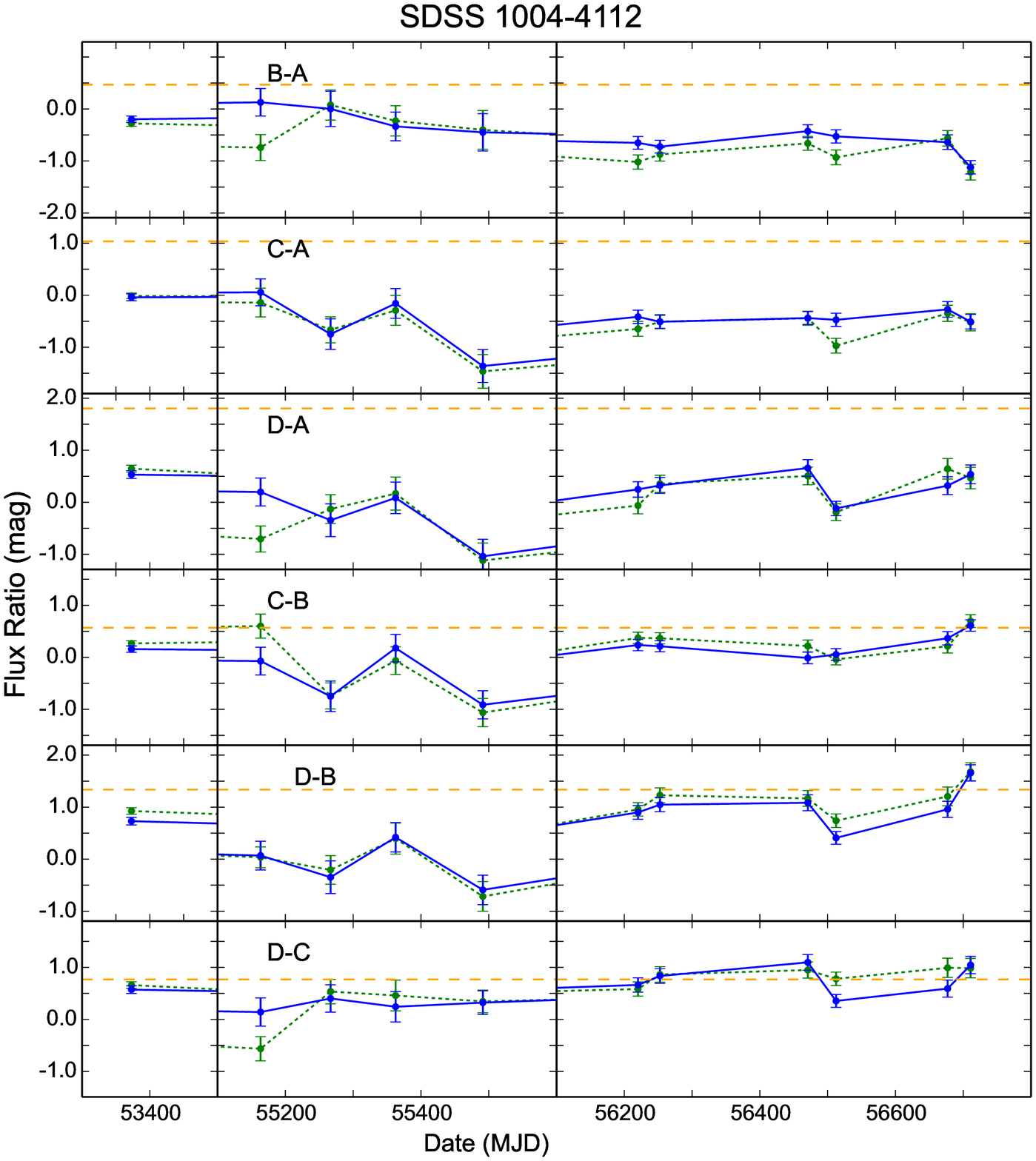}
\caption{Flux ratios for SDSS 1004+4112 on a magnitude scale. Continuous blue (dashed green) curves show the hard (soft) emission. The orange dashed horizontal lines represent the baseline ratios.\label{fig:1004}}
\end{figure}

\begin{figure}
\epsscale{.80}
\plotone{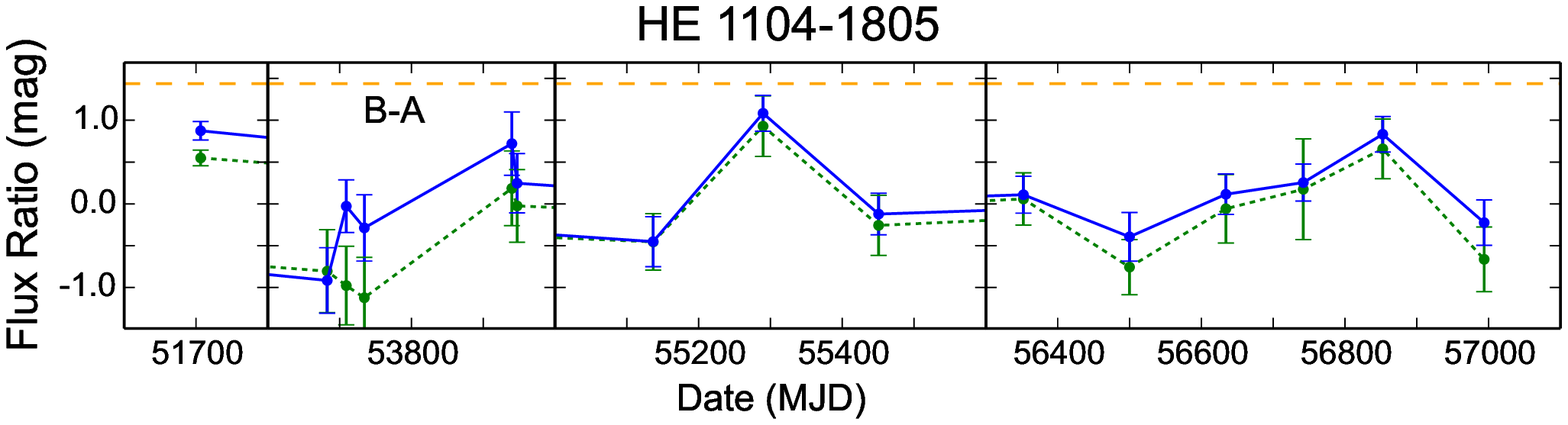}
\caption{Flux ratios for HE 1104$-$1805 on a magnitude scale. Continuous blue (dashed green) curves show the hard (soft) emission. The orange dashed horizontal lines represent the baseline ratios.\label{fig:1104}}
\end{figure}

\begin{figure}
\epsscale{.80}
\plotone{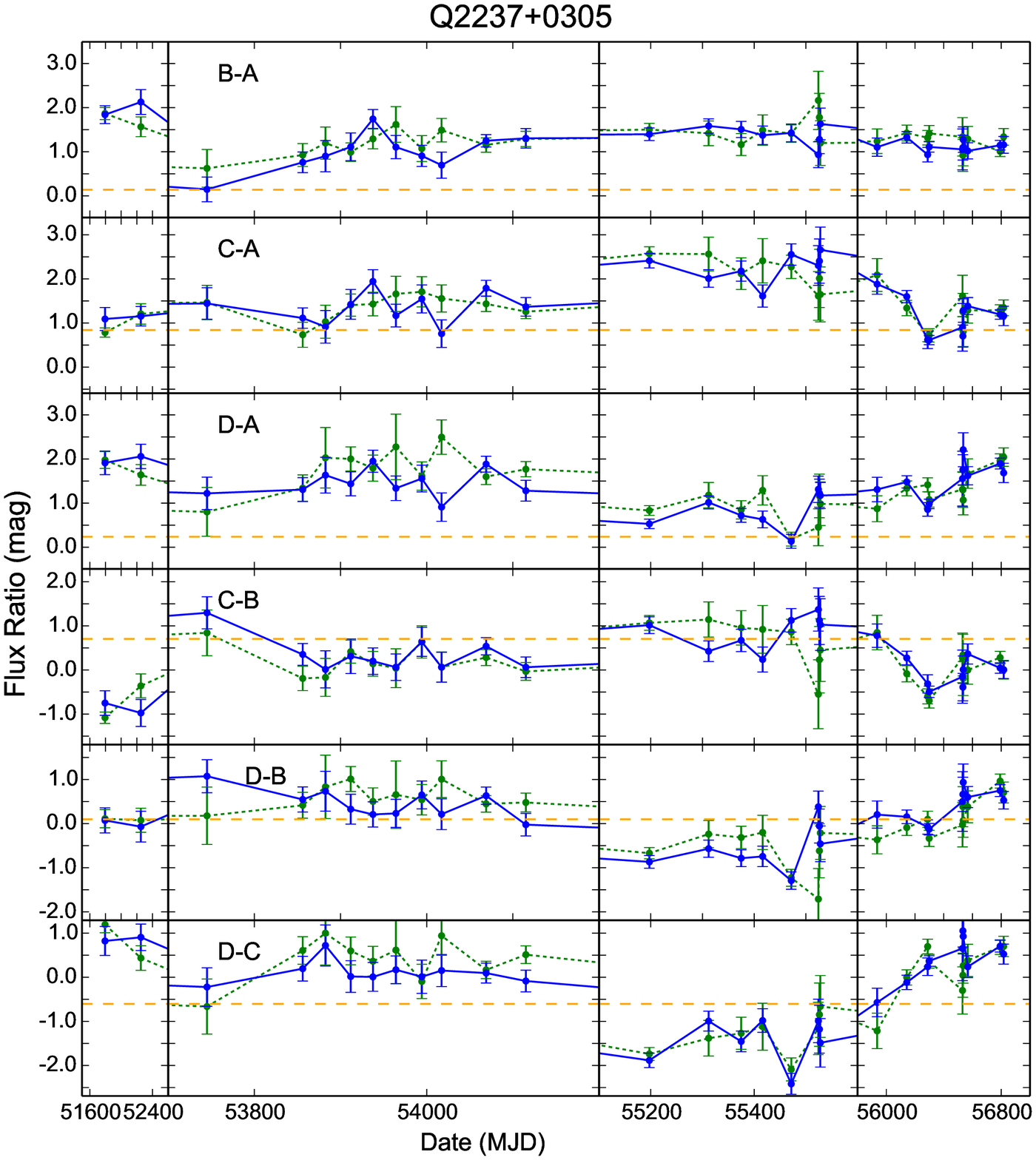}
\caption{Flux ratios for Q 2237+0305 on a magnitude scale. Continuous blue (dashed green) curves show the hard (soft) emission. The orange dashed horizontal lines represent the baseline ratios.\label{fig:2237}}
\end{figure}

We want to compare the microlensed flux ratios between images with the intrinsic flux ratios that are not affected by microlensing (baseline ratios). After correction for any absorption, the baseline flux ratios are primarily determined by the smooth potential of the lens galaxy, although they may be perturbed by substructures in the lens galaxy such as satellite halos \citep{KochankekDalal2004,Zackrisson2010}. Ideally, the baseline ratios are measured at wavelengths where the quasar emitting region is much larger than the Einstein radius (several light days for lensed quasars) and therefore not sensitive to microlensing. This is generally true of radio, rest-frame mid-IR and narrow line emission. When that is not possible, baseline ratios can be approximated using the macro magnifications from lens models. We adopt the latter approach, with baseline ratios derived from macro lens models. The values adopted are shown as horizontal lines in Figures~\ref{fig:0158}--\ref{fig:2237}.

The measured flux $f_{ij}$ (count rate) of the i-th image at the j-th epoch
\begin{equation} \label{eq:fluxratios}
 f_{ij} = s_{j} \cdot \mu_{i} \cdot \xi_{ij}, 
\end{equation}
is the source flux $s_{j}$ magnified by a combination of macro-lens and microlensing magnifications $\mu_{j}$ and $\xi_{ij}$, respectively. The flux ratio between two images $A$ and $B$ is
\begin{equation} \label{eq:equation10}
 \frac{f_{Bj}}{f_{Aj}} = \frac{ s_{Bj} }{ s_{Aj} } \cdot \frac{ \mu_{B} }{ \mu_{A} } \cdot \frac{ \xi_{Bj} }{ \xi_{Aj} }. 
\end{equation}
We are interested in $\xi_{Bj} / \xi_{Aj}$, but we measure the fluxes ${f_{Bj}}$ and ${f_{Aj}}$. An additional complication is that the source flux ratio is really
\begin{equation}
\frac{ s_{Bj} } { s_{Aj} } = \frac{ s(t) }{ s(t + \delta t_{AB}) }, 
\end{equation}
which includes a propagation time delay $\delta t_{AB}$ \citep{Refsdal1964,Cooke1975} that cannot be easily removed from the sparsely sampled X-ray light curves as is done in optical studies \citep[e.g., ][]{Kochanek2006,Tewes2013}. If $\delta t_{AB}$ is much smaller than the typical time scale of intrinsic variability, the effects of the time delay become unimportant. This is certainly the case of Q 2237+0305 where the time delays are constrained to be $< 1$ day \citep{Dai2003}. The other extreme in our sample is SDSS 1004+4112, where the delays are as long as several years \citep{Fohlmeister2008}. In these cases, the time delays combined with intrinsic source variability add ``noise'' to the light curves that can be interpreted as additional microlensing variability. We will follow the usual procedure \citep[e.g., ][]{Schechter2014} and assume that $s_{Bj}/s_{Aj} \approx 1$. This strategy is safest for image pairs with shorter lens delays. Using this assumption, Equation \ref{eq:equation10} becomes
\begin{equation} \label{eq:equation40}
\chi_{BA}(t_{j}) = \frac{ \xi_{Bj} }{ \xi_{Aj} } = \left [ \frac{f_{Bj}}{f_{Aj}} \right ] \cdot \left [ \frac{ \mu_{B} }{ \mu_{A} } \right ]^{-1} .
\end{equation}
This \textit{microlensing amplitude} should not be confused with simple flux ratios which include no corrections for the macro lens magnifications. The microlensing magnification ratio can also be expressed in magnitudes to facilitate comparison with optical microlensing studies where the image fluxes and macro magnification ratios are now expressed in magnitudes,
\begin{equation}
m_{B} - m_{A} = -2.5 \log (f_{B}/f_{A})
\end{equation}
\begin{equation}
m_{B}^{0} - m_{A}^{0} = -2.5 \log (\mu_{B}/\mu_{A})
\end{equation}
\begin{equation}\label{eq:equation41}
\Delta m_{AB} = (m_{B} - m_{A}) - (m_{B}^{0} - m_{A}^{0}).
\end{equation}
Table~\ref{tab:kappasgammas} summarizes the values we adopt for the macro magnifications, where the total magnification $\mu_{i}$ is derived from the estimated convergence $\kappa_{i}$ and shear $\gamma_{i}$ at each image position according to $\mu_{i} = [ (1- \kappa_{i})^{2} - \gamma_{i}^2 ]^{-1}$ \citep[see, e.g., ][]{Narayan1996}. These estimates of $\kappa_{i}$ and $\gamma$ are obtained by fitting a model for the overall potential of the lens galaxy to the lens data. We fit a singular isothermal ellipsoid with external shear (SIE+g) to both QJ~0158$-$4325 and HE 1104$-$1805 using \textit{Lensmodel} \citep{Keeton2001}. For SDSS 1004+4112, we used the cluster mass model for this lens by \cite{Oguri2010}, ($\kappa, \gamma$ values given in private communication). For the rest of the objects, we used SIE+g results from the literature as listed in Table~\ref{tab:kappasgammas}.

We also need the surface density of stars $\kappa_{*}$ relative to the total surface density $\kappa$. We combined the astrometry of each lens with the compilation of lens galaxy effective radii in \cite{Oguri2014} to estimate the ratio $R/R_{eff}$ between the radial distance of each image from the lens center and the effective radius of the lens. We then used the best fit model for $\kappa_{*}/\kappa$ from \cite{Oguri2014} to estimate the stellar surface density, except in the case of SDSS 1004+4112. The lens SDSS 1004+4112 is a special case because it is a cluster lens, where the member galaxies are further from the images than in single-galaxy lenses. We adopted an arbitrarily low value $\kappa_{*}/\kappa=0.03$ as pleausible estimate of the low optical depth associted with intracluster stars. Q 2237+0305 is also a special case because the images are seen through a galactic bulge, and here $\kappa_{*}/\kappa \simeq 0.8$ because the images are seen through a galactic bulge. The average value of $\kappa_{*}/\kappa$ is in reasonable agreement with previous estimates \citep{Jorge2015}. The results are presented in Table~\ref{tab:kappasgammas}.

There are two ways in which microlensing effects may manifest themselves. One is through the time variability in the flux ratios, which is independent of the baseline flux ratios but can be affected by intrinsic variability modulated by time delays. The second is if the X-Ray flux ratios differ from the estimated base line ratios. The difference allows a quantitative measurement of microlensing but it is only reliable to the extent that the baseline flux ratios are accurate. We summarize the differential microlensing in Figures~\ref{fig:0158}--\ref{fig:2237}, where the flux ratios for the hard and soft X-ray bands are compared and the baseline ratios are shown as horizontal lines. The distance between the data points and the horizontal lines represent the differential microlensing. The figures show a complex pattern of time variability attributable to microlensing and, to so some extent, to noise introduced by time variabiliy.

\subsection{An Independent Test for Energy-Dependent Microlensing}
The X-ray emission regions of lensed quasars appear to be more compact than the disk emission seen at ultraviolet or optical wavelengths because the X-ray microlensing amplitude is consistently higher \citep[e.g., ][]{Morgan2008,Chartas2009,Dai2010,Pooley2012,Mosquera2013,Schechter2014}. This has been successfully used to map the size of the accretion disk at different wavelengths \citep{Poindexter2008, Mediavilla2011a, Blackburne2011}, but X-ray microlensing analyses to date have found that the hard and soft band X-ray emission regions are of similar size \citep{Morgan2012,Mosquera2013,Blackburne2014,Blackburne2015}. Here we want to test if the new data remain consistent with the null hypothesis that the microlensing amplitude is the same for both bands. In the absence of any band-dependent difference in the microlensing amplitude, the data should be consistent with the soft and hard bands having a common microlensing magnification ratio $\xi_{ij}$ (see Equation \ref{eq:equation40}). We can test for this by optimising the statistic
\begin{equation}
\chi^2=\sum_{i,j}^{}  {
\left[
  \frac   { \left( f^{\rm hard}_{ij}-(s^{\rm hard}_j \cdot \mu_i \cdot \xi_{ij}) \right) ^2}   { \left( \sigma^{\rm hard}_{ij} \right) ^2}
+ \frac   { \left( f^{\rm soft}_{ij}-(s^{\rm soft}_j \cdot \mu_i \cdot \xi_{ij}) \right) ^2}   { \left( \sigma^{\rm soft}_{ij} \right) ^2}
\right]
                      },
\end{equation}
with respect to the source fluxes ($s^{\rm soft}_j$, and $s^{\rm hard}_j$) and the microlensing magnification ratio $\xi_{ij}$. If the source sizes are comparable, we should obtain a good fit using a single value for the microlensing magnification. The macrolens magnifications $\mu_{i}$ are fixed to the estimates from Table~\ref{tab:kappasgammas} and should be the same for all energy bands. The subscripts $i$ and $j$ refer to image and epoch respectively.

We can apply this test to the $4$ image lenses, finding reduced $\chi^2$ values of $0.79, 0.94, 1.5,$ and $1.4$ for HE 0435$-$1223, SDSS 0924+0219, SDSS 1004+4112, and Q 2237+0305 respectively, given $23, 15, 25,$ and $63$ degrees of freedom. The p-values are $0.75, 0.52, 0.05,$ and $0.01$. This implies the existence of energy dependent microlensing in Q 2237+0305 and, to a lesser extent, in SDSS 1004+4112. It must be noted that a high $\chi^2$ value can be the result of consistently wider amplitudes of either the hard or soft microlensing ratios. However, it may also happen as a result of non related signals of similar amplitude. Therefore a connection between this results and an intuitive interpretation of Figures~\ref{fig:0158}--\ref{fig:2237} is not straightforward.

\subsection{Root-Mean-Square of Microlensing Variability}

Microlensing flux variations occur as the stars in the lens galaxies move relative to the background source, producing a complex variability pattern. Here we want to explore the root-mean-square (RMS) of the microlensing amplitude as an observable that can potentially be related numerically to the physical properties of the lens system. This is a reasonable assumption since a small source crossing a region with high density of caustics will show larger flux variations. If $\chi_{BA}(t_i)$ is the microlensing amplitude (as defined in Equation~\ref{eq:equation40}) of a certain image pair at epoch $t_i$, then we define
\begin{equation}
\overline{\chi}_{BA} = \frac {1}{N}\sum _{i=1}^{N}\chi_{BA}(t_i)
\end{equation}
\begin{equation}
\chi_{BA}^{[RMS]} = \sqrt {{\frac {1}{N-1}}\sum _{i=1}^{N}(\chi_{BA}(t_i)-\overline{\chi}_{BA} )^{2}}
\end{equation}
Table~\ref{tab:mean_rms_table} lists the these two statistics for the full, soft, and hard bands along with their uncertainties. To quantify the significance of the differences between the soft and hard band RMS values, Table~\ref{tab:mean_rms_table} also shows the p-values for rejecting the null hypothesis of identical distributions computed with a Welch two-tail test. The p-values suggest a distinct physical origin for the soft and hard band at a significance greater than $2\sigma$ confidence level in $8$ out of $20$ image pairs, and in $12$ out of $20$ pairs at a significance level greater than $1\sigma$. However, the sign of the difference between the hard and the soft band RMS is not consistent accross the sample, i.e. this difference does not always show the same sign across the three components of each quadruple quasar. This suggests the need to compare the results with an analysis of their a priori probabilities (Guerras et al. 2017, in preparation).

\begin{figure}
\epsscale{1.00}
\plotone{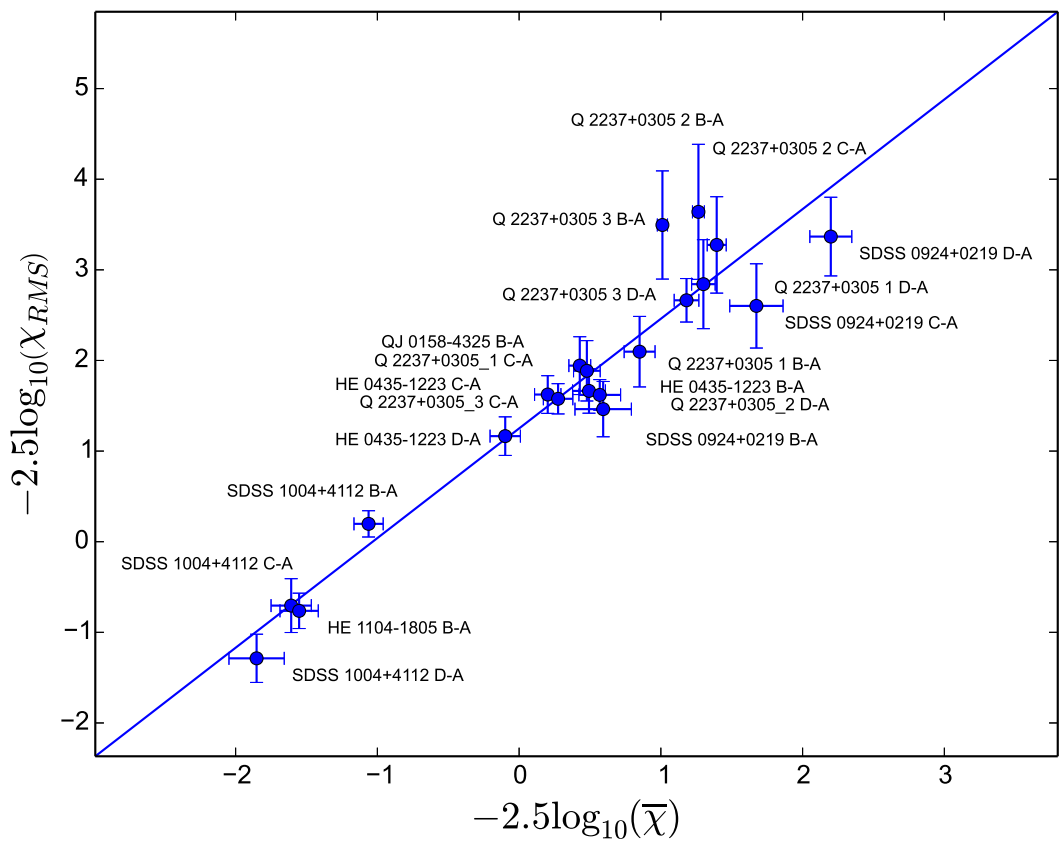}
\caption{The empirical relationship between the RMS (root-mean-square) and mean of the full band microlensing amplitude (defined in Eq. \ref{eq:equation40}). The best linear fit in log space is shown by a solid line. Given the extraordinarily long campaign for Q 2237+0305, its data have been broken in $3$ chunks of approximately equal size to roughly match the other objects. \label{fig:rms_linear}} 
\end{figure}

Figure~\ref{fig:rms_linear} shows that the RMS $\chi^{[RMS]}$ and mean $\overline{\chi}$ microlensing signal are correlated. We find a Pearson correlation coefficient of $R = 0.96^{+0.03}_{-0.06}$ ($95\%$ CL) and a best fit correlation of
\begin{equation}
 \log_{10} \left(\chi^{[RMS]}\right) = (1.21 \pm 0.08) \cdot \log_{10}\big(\overline{\chi}\big) - (0.50 \pm 0.03) .
\end{equation}
This relationship suggests that the RMS may be a useful observable to better constrain the physical properties of lensed quasars. We will explore these issues further in Guerras et al.\ (2017, in preparation).

\section{Source Size Estimates}\label{sec:source_size_estimates}


Next we are interested in analysing the departures of the flux ratios from the base line ratios to determine the source size. The first step is a quantitative characterisation of such departures. When the observation campaigns cover short periods as compared with the microlensing variability timescales, a common approach is to assume that the span of the data is too short to observe microlensing variability. Several epochs are then averaged \citep[e.g.,][]{Pooley2012,Jorge2015,Munoz2016} and the resulting averages are compared with the numerical predictions for one single epoch per image pair \citep[e.g.,][]{Mediavilla2009,Blackburne2011,Guerras2013a}.

There are two characteristic microlensing variability timescales. One is the Einstein radius (Equation~\ref{eq:AE}) crossing time, which is generally quite long, and a second, shorter timescale associated with the source crossing time. Table~\ref{tab:summaryofdata} summarizes estimates for both timescales from \cite{MosqueraKochanek2011}. These estimates strongly suggest that the comparison of simple averages of the light curves against single-epoch model predictions will be suboptimal here because our present data have time spans long enough that we should expect microlensing variability. By collapsing the light curves into average values, we could lose information because the behaviour of the average signal may not be well-modelled by single-epoch predictions.

\begin{figure}
\epsscale{1.00}
\plotone{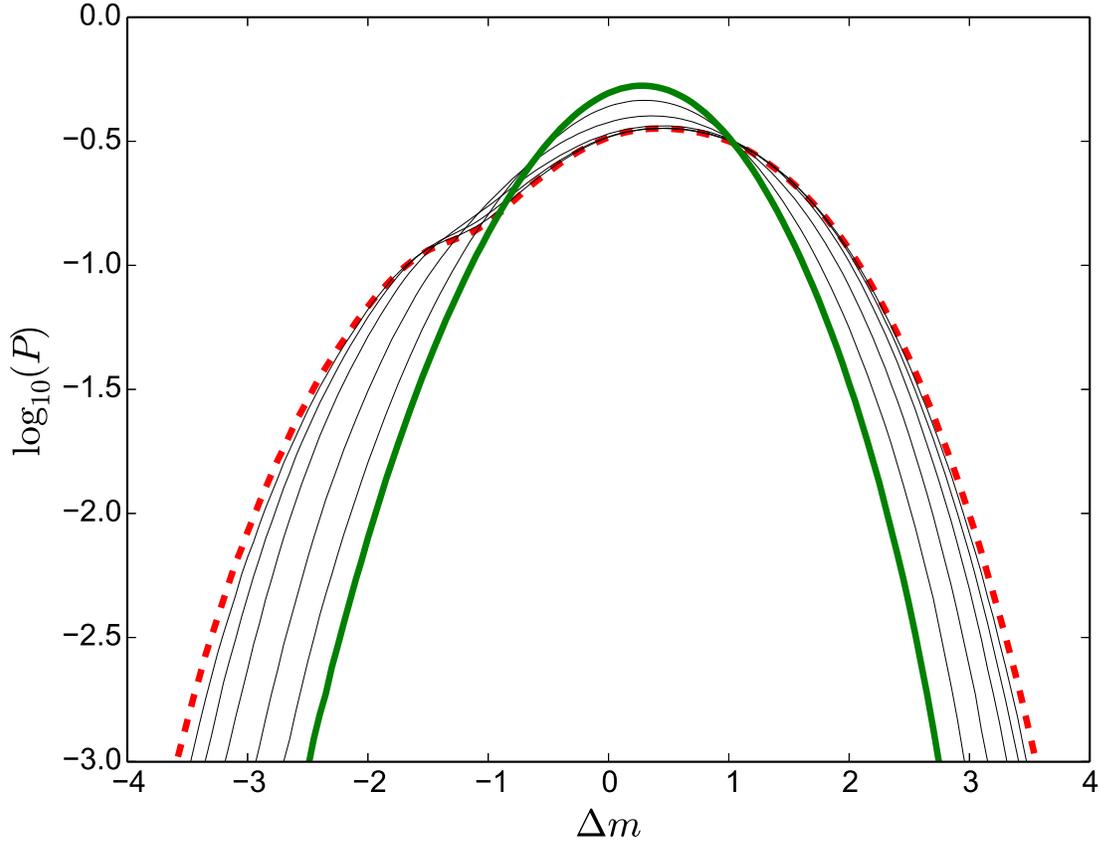}
\caption{Probability Density Functions of differential microlensing between image $C$ and $A$ of Q 2237+0305 for a source with a half-light radius of $0.5$ light-day. The solid, green thick line shows the distribution function obtained from $10^8$ simulated observation campaigns spanning $1.7$ $R_E$ each with $30$ observations, which roughly matches our observations. The dashed, red thick line shows the analogous distribution from $10^8$ simple single-epoch observations. The results for intermediate track lengths of $0.05$, $0.10$, $0.20$, $0.50$, and $1.00$ $R_E$ (thin solid lines) are also shown. The longer the averaged light curves, the higher the departure from a single-epoch probability distribution. \label{fig:histogramas}}

\end{figure}

To explore the impact of averaging long observation campaigns on the probability of differential microlensing magnification, we can compare computer-generated probability distributions (details on their generation are given in Section \ref{subsec:magmaps}) of single-epoch differential microlensing with analogous simulations where the predicted quantity is the average of differential microlensing along randomly orientated tracks whose length in Einstein Radii and number of observations match those in our data sample based on the scales in Table~\ref{tab:summaryofdata}. Figure \ref{fig:histogramas} illustrates this for Q~2237+0305 (C-A). Based on the $13.6$ year timespan of our data, and the estimate given by \cite{MosqueraKochanek2011} of the time it takes the source to cross the Einstein radius in this system, the source has moved roughly $1.7$ $R_E$. We computed the mean microlensing signal observed by averaging over 30 evenly spaced epochs where the source moves from $0$ $R_e$ (i.e. a single epoch) up to $1.7$ $R_E$. The distribution of mean magnifications begins to narrow relatively quickly, particularly in the wings of the distributions. Thus, not taking into account the time averaging will lead to the derivation of an overly large source size because the source size must compensate for the neglected temporal smoothing. One compromise to address this problem is to only average epochs separated by shorter timescales \citep[as done by e.g.,][]{Munoz2016}. We will instead use numerical models that take the length of each observation campaign into account.

We will derive a probability distribution for the half-light radius of the source, using all 4 or 2 images simultaneously. Given one object, the flux (count rate) of image $\alpha$ expressed in magnitudes at epoch $t_i$ is
\begin{equation}
 m^{obs}_{\alpha}(t_i) = m_{0}(t_i) + \mu_{\alpha} + \xi_{\alpha}(t_i)
\end{equation}
where $m_{0}(t_i)$ is the intrinsic magnitude of the source at epoch $t_i$, $\mu_{\alpha}$ is the macrolens magnification of image $\alpha$ and $\xi_{\alpha}(t_i)$ is the microlensing magnification of image $\alpha$ at epoch $t_i$. Following \cite{Kochanek2004} we first eliminate the intrinsic magnitude of the source by optimising a source model $m_{0}(t_i)$ simultaneously to fit all the images:
\begin{equation}\label{eq:sourcefit}
\chi^2(t_i)=\sum_{\alpha} \Big ( \frac{m^{obs}_{\alpha}(t_i) - [m_{0}(t_i) + \mu_{\alpha} + \xi_{\alpha}(t_i)]}{\sigma_{\alpha}} \Big )^{2}
\end{equation}
After substituting the best source model $m_{0}(t_i)$ the statistic in Eq.~\ref{eq:sourcefit} reduces to
\begin{equation}
\chi^2(t_i)=\sum_{\alpha} \sum_{\beta < \alpha} \Big ( \frac{ \left( m^{obs}_{\alpha}(t_i) - [\mu_{\alpha} + \xi_{\alpha}(t_i)] \right) - \left( m^{obs}_{\beta}(t_i) - [\mu_{\beta} + \xi_{\beta}(t_i)] \right)  }{\sigma_{\alpha \beta}} \Big )^{2}
\end{equation}
where the errors $\sigma_{\alpha \beta}$ are computed according to Equation $(7)$ in \cite{Kochanek2004}.
This expression can be rearranged as
\begin{equation}
\chi^2(t_i)=\sum_{\alpha} \sum_{\beta < \alpha} \Big ( \frac{ \Delta m^{obs}_{\beta \alpha} - \Delta m_{\beta \alpha} }{\sigma_{\beta \alpha}} \Big )^{2}
\end{equation}
where $\Delta m^{obs}_{\beta \alpha}=\left( m^{obs}_{\beta}(t_i) - \mu_{\beta} \right)-\left( m^{obs}_{\alpha}(t_i) - \mu_{\alpha} \right)$ is the observed microlensing magnification as defined in Equation \ref{eq:equation41} and $\Delta m_{\beta \alpha} = \xi_{\beta}(t_i) - \xi_{\alpha}(t_i)$ is the microlensing magnification predicted by a numerical model. We explain the generation of the $\Delta m_{\beta \alpha}$ in Section \ref{subsec:magmaps}.

Given the results for one trial, the likelihood of the source size for each epoch $i$ can be obtained by adding the likelihoods of a high number $N$ of trials,
\begin{equation}
L_{i}(r_s) \propto \sum^{N}\exp{\Big( -\frac{1}{2}\chi_i^2 \Big)}.
\end{equation}
Following other studies, e.g. \cite{Munoz2016}, \cite{Jorge2015}, we will collapse the light curves by averaging them \footnote{Although there are small differences among epochs, we set the uncertainty as the average of the measurement errors for each image.} to single time independent values $m^{obs}_{\alpha}$. The difference here will be in the model we use to generate the predicted distribution of the differential microlensing amplitudes $p\big(\Delta m_{\beta \alpha} \big| r_{s} \big)$. Rather than simply using the results for a single point, we will use averages obtained from random sets of data tracks that emulate the length and time sequence of the measurements available for each lens. Once we get a time-independent likelihood function $L(r_s)$ for each lensed quasar, the joint probability density function is obtained as the normalised product of the individual likelihood functions,
\begin{equation}\label{eq:definelikelihood} 
 P(r_{s}) \propto \prod_{j} L_{j} (r_{s}).
\end{equation}

\subsection{Computer-generated probability distributions}\label{subsec:magmaps}
To obtain the probability distributions $p \big(\Delta m_{\beta \alpha} \big| r_{s} \big)$ for each image pair and source size, we generated magnification patterns based on the $3$ local parameters for each image given in Table \ref{tab:kappasgammas} (the local surface mass density $\kappa$, the shear $\gamma$, and the fraction of the local surface density in stars $\kappa_{*}/\kappa$). The stars are assigned a fixed mass of $M_{*}=0.3 M_{\odot}$, since it has been shown that microlensing statistics depend little on the stellar mass function \citep[e.g.,][]{Wambsganss1992, Mediavilla2015}. The size estimates can be easily re-scaled to a different mean mass as $r_{s} \propto \sqrt{M_{*}}$.

To model the effect of the finite source size, we convolve the maps with a Gaussian kernel $I(r) \propto \exp[-r^2/(2r_{s}^{2})]$. For comparisons to other profiles, the half-light radius $r_{1/2}=1.18$ $r_{s}$ should be used since estimates of $r_{1/2}$ are insensitive to profile changes \citep{Mortonson2005}. The size of the X-ray corona is expected to be proportional to the mass of the central black hole \citep{Mosquera2013, Jorge2015}, so a natural choice for the source scaling is in units of gravitational radii $R_{g} = GM_{BH}/c^{2}$ based on the estimates given in Table \ref{tab:summaryofdata}. For this case we used a grid where $R_{s}/R_{g}=e^{0.15n}$ with $n=0,1, 2, ...,35$. We used the $0.025$ light-day/pix maps except in those cases where $r_{s}$ would be below $1$ pixel in size, where we switched to the $0.006$ light-day/pix maps. Our size estimates can be rescaled to other choices of the mean stellar mass as $r_{s} \propto \sqrt{M_{*}}$.

Each $p \big(\Delta m_{\beta \alpha} \big| r_{s} \big)$ is generated as the normalized histogram of $10^8$ trials on the maps for the corresponding lens, images $\alpha$ and $\beta$, and source size $r_s$. Each trial consists on a randomly oriented track whose length and time sequence corresponds to the real observation campaign of the object, placed on a random position on each map. The simulated light curves are then averaged identically as with the observational light curves. The track length for each object in Einstein radius units is obtained from the ratio of the time span covered by its observational campaign to the Einstein radius crossing time estimates given by \cite{MosqueraKochanek2011} and summarized Table \ref{tab:summaryofdata}.

\subsection{Results}

\begin{figure}
\epsscale{1.00}
\plotone{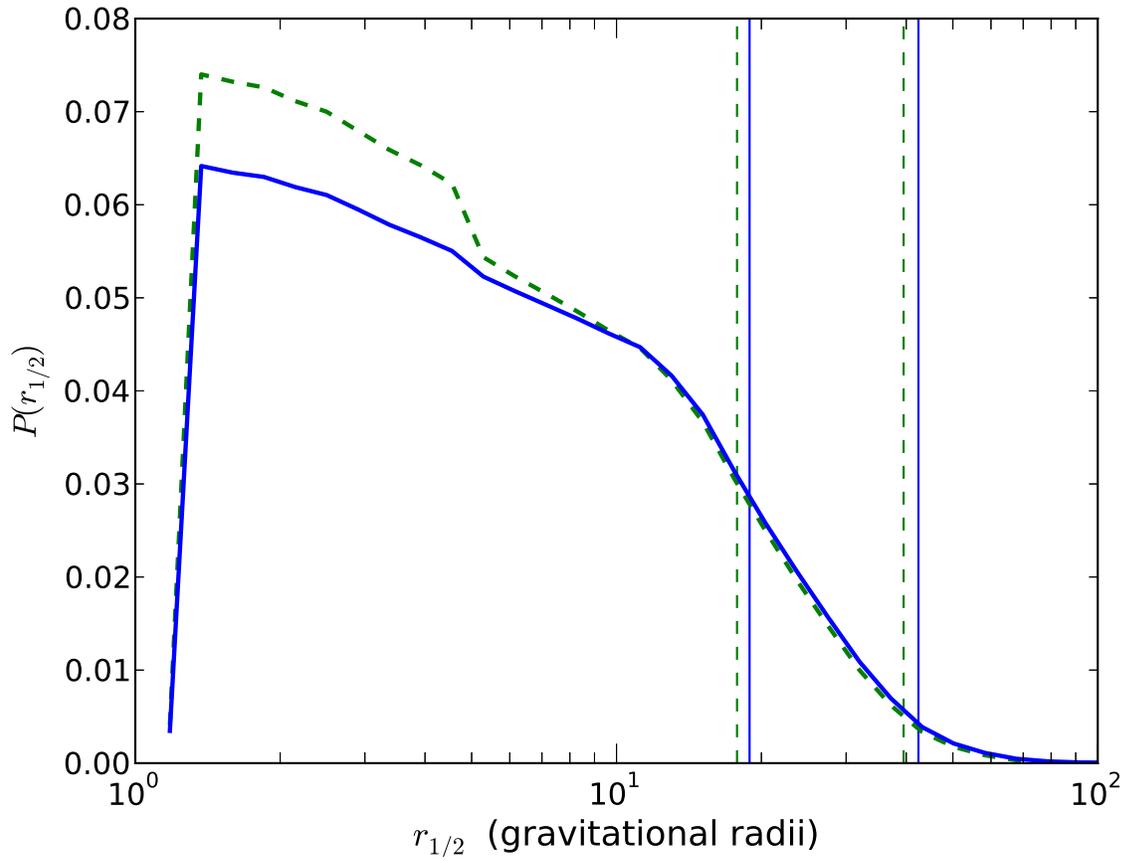}
\caption{Joint probability distribution for the average half-light radius. The hard (soft) X-ray band result is shown by the continuous blue (dashed green) curve. The vertical lines show the corresponding $68\%$ and $95\%$ one-sided probability upper limits.  \label{fig:joint_likelihood_curves}}
\end{figure}

We followed the same procedure for the soft and hard bands, obtaining the joint probability distributions for $r_s$ shown in Figure \ref{fig:joint_likelihood_curves}. The expected values are $15.1 \pm 12.6$ ($16.3 \pm 14.7$) gravitational radii for the half-light radius of the soft (hard) band. For the average black hole mass in our sample, this translates into $0.42 \pm 0.35$ ($0.46 \pm 0.41$) light-days for the soft (hard) band. However, the probability distributions peak near $1 R_{g}$, so only the upper limits are meaningful in practice. We get \textit{upper limits} of $17.8$ ($18.9$) and $39.5$ ($42.4$) gravitational radii for the the soft (hard) band at $68\%$, and $95\%$ one-sided confidence limits, respectively, or $0.50$ ($0.53$) and $1.1$ ($1.2$) light-days for the the soft (hard) band at $68\%$, and $95\%$ one-sided confidence limits, respectively. The results are shown in Table \ref{tab:softandhardresults_gr}, where the $99\%$ confidence values are also included.

\begin{figure}
\epsscale{1.00}
\plotone{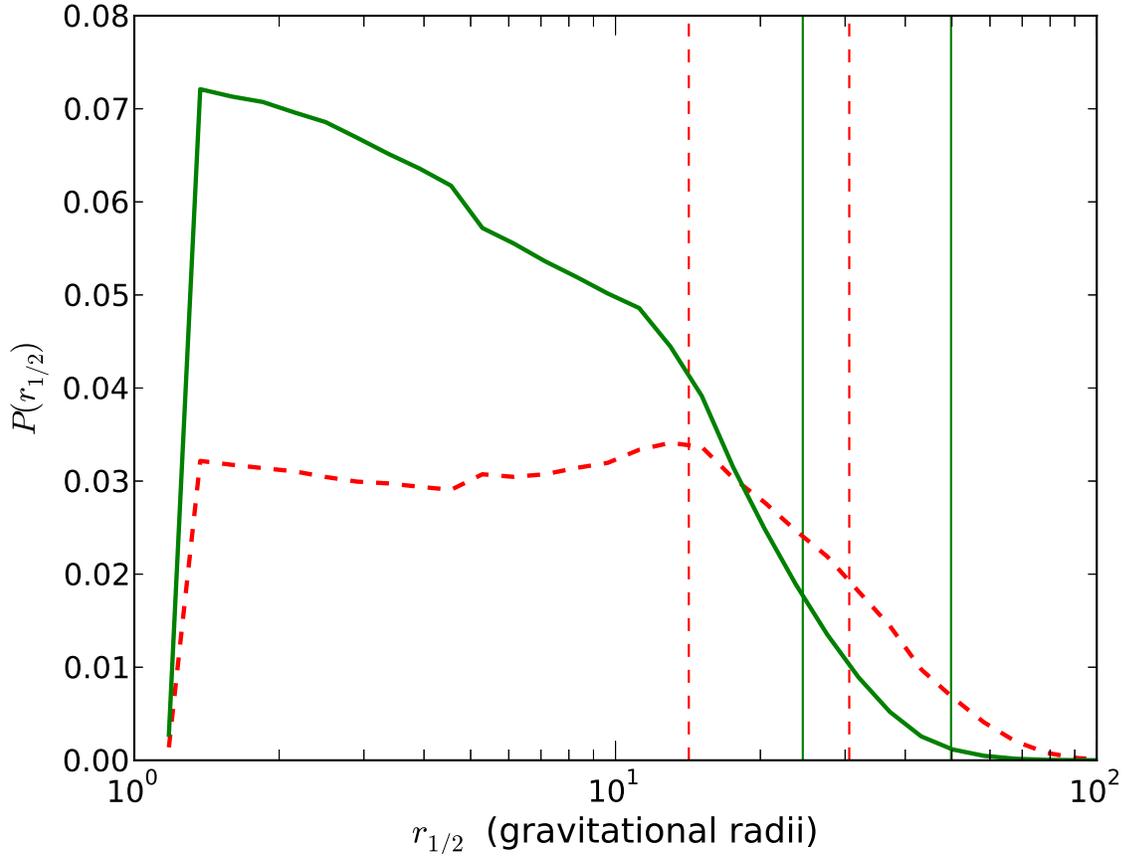}
\caption{Joint probability distribution for the average half-light radius of the full X-ray band (continuous green). Using single-epoch histograms neglects the effect of temporal smoothing, and the resulting distribution (dashed red) yields overestimated source sizes. The vertical lines show the corresponding $68\%$ and $95\%$ one-sided probability upper limits. \label{fig:joint_likelihood_tracks_vs_points}}
\end{figure}

When the calculations are performed on the full band, the expected value for the half-light radius is $14.1 \pm 10.9$ gravitational radii ($0.40 \pm 0.31$ light-days) and the $68\%$ probability upper limit on the half-light radius is $16.8$ gravitational radii ($0.47$ light-days). For comparison purposes, we repeated the calculation ignoring the effects of temporal smoothing by using single-epoch histograms for $p\big(\Delta m_{\beta \alpha} \big| r_{s} \big)$ (Figure \ref{fig:joint_likelihood_tracks_vs_points}). For this case we obtained a expected value of $24.0 \pm 17.7$, and a $68\%$ probability upper limit of $28.9$ gravitational radii. This illustrates the impact of neglecting the length of the observation campaigns. Treating our time averaged data as single-epoch data overestimates the source size by a factor $1.7$


\section{Discussion}\label{sec:discussion}

We have measured full, soft and hard band X-ray light curves for $6$ lensed quasars to look for microlensing by comparing the observed flux ratios with the ratios predicted by macro lens modeling. We have tested for energy-dependent variability in several ways: a $\chi^2$ fit to the light curves of quadruple lenses, a comparison between the microlensing amplitude RMS of the soft and hard bands, and estimates of the average source size in the full, hard and soft X-ray energy bands.

Our $\chi^2$ test for energy-dependent microlensing shows a lack of correlation between soft and hard band in 2 of the 4-image lenses. This can be explained by a size difference between emitting regions, but also by a lack of correlation in the time domain. The RMS of the microlensing variability between the hard and soft bands is significantly different for a number of image pairs, but the sign of the difference varies and shows no consistent pattern. If a higher RMS is interpreted as arising from a more compact hard X-ray source, then this picture is consistent with the recent review of a sample of $8$ lensed quasars by \cite{Chartas2016}, where for some objects the hard X-ray emission regions seems to be more compact than the soft and in others the soft appears to be smaller. However, a physical interpretation of a higher RMS level of microlensing variability in one band might not be straightforward, as suggested by the inconsistencies shown in Table~\ref{tab:mean_rms_table} among image pairs of the same quadruple objects. An analysis on this question is in preparation (Guerras et al. 2017). 

Our estimates of the average source size indicate that any size difference between the hard and soft emitting regions must be modest. This is in good agreement with the general picture that emerges from fully time-dependent studies of individual objects. \cite{Blackburne2015} found the same upper limit for the size of the hard and soft X-ray emitting regions in HE 1104$-$1805, as did \cite{Morgan2012} for QJ 0158$-$4325. \cite{Mosquera2013} could find only ``weak evidence'' that the hard X-ray emitting region in Q 2237+030 was more compact than the soft X-ray emitting region, and in HE 0435$-$1223 \cite{Blackburne2014} found no evidence for a size difference. The physical structure of what is believed to be a hot corona responsible for the X-ray continuum in quasars is poorly understood, and there are other astrophysical examples where hotter does not necessary equal smaller (e.g., the Solar corona). There is even some evidence \citep{Chartas2012} suggesting that the soft emitting region could be more compact at least in one case.

The X-ray light curves span time intervals comparable to typical microlensing time scales, Treating them as single-epoch observations would result in overestimated source sizes. One way to address this problem is with fully time dependent calculations \citep{Kochanek2004}, but these are very computationally expensive. Here we introduce a simple approximation which at least avoids the bias of the single epoch method. This approximation essentially consists of introducing probability distribution modelled from time averaged tracks across the magnification patterns instead of isolated data points. The size estimates are consistently smaller using this approach. If we do not include the effects of time averaging, the source size estimate increases by a factor of $1.7$, which is a significant bias in the size estimate.

We also introduced a radial model for the microlensing optical depth based upon the best fit to real data given by \cite{Oguri2014}. This model is simpler than the de Vaucouleurs stellar distribution plus NFW dark-matter halo used in previous time-dependent studies on individual objects \citep[e.g.,][]{Morgan2008, Dai2010}, yet it is an improvement with respect to using a uniform value in previous single-epoch studies over a heterogeneous sample of quasars \citep[e.g.,][]{Guerras2013b, Jorge2015}, where it is desirable to do the least possible assumptions on the lens galaxies to give a uniform treatment to all objects in the sample.

We also find a functional relationship between the RMS and the average value of the microlensing amplitude. This suggest that both observables carry physical information (e.g., about the quasar source size or the optical depth in galactic halos), and a more detailed analysis of this correlation is in preparation (Guerras et al. 2017). These two observables could be used complementarily to constraint physical properties from microlensing variability for a better understanding of lensed quasars.

    
\section{Acknowledgements}
We thank Jorge Jim\'enez-Vicente for useful discussions about the magnification patterns. We thank Evencio Mediavilla for providing an implementation of the Inverse Polygon Mapping code, and Masamune Oguri for providing the convergence and shear values of SDSS 1004+4112 calculated from his mass model. We would like to thank the anonymous reviewer for the useful comments and suggestions provided.

Support for this work was provided by the National Aeronautics and Space Administration through {\emph{Chandra}} Award Number GO0- 11121A/B/C/D, GO1-12139A/B/C, GO2-13132A/B/C, and G03-14110A/B/C issued by the {\emph{Chandra}} X-ray Observatory Center, which is operated by the Smithsonian Astrophysical Observatory for and on behalf of the National Aeronautics Space Administration under contract NAS8-03060. XD acknowledges NASA ADAP program NNX15AF04G and NSF grant AST–1413056. CSK is supported by NSF grant AST-1515876.


\clearpage


\clearpage

\begin{deluxetable}{lcccccccc}
\tabletypesize{\scriptsize}
\tablecaption{Lens Data\label{tab:summaryofdata}}
\tablewidth{0pt}
\tablehead{
\colhead{Object} & \colhead{$z_{s}$} & \colhead{$z_{l}$} & \multicolumn{1}{c}{$R_{E}$} & \colhead{$t_E$} & \colhead{$t_s$} & \colhead{$\Delta t_{obs}$} &  \colhead{$\Delta t_{obs}/t_E$}  &\colhead{$M_{BH}$}         \\
                 &                   &                   & (light-days)                & (years)         & (years)         & (years)                    &  (Einstein radii)                &($\times 10^{9}$ $M_{\odot}$)
}

\startdata

QJ 0158$-$4325  &  1.29   &  0.317   &  7.434 & 18.0 & 0.86  &  4.6   &  0.26  &  $0.16$ $ $  $ $(MgII)\\
HE 0435$-$1223  &  1.689  &  0.46    &  7.986 & 18.3 & 0.47  &  7.3   &  0.40  &  $0.50$ $ $  $ $(CIV) \\
SDSS 0924+0219  &  1.524  &  0.39    &  7.790 & 20.4 & 0.39  &  5.6   &  0.27  &  $0.11$ $ $  $ $(MgII)\\
SDSS 1004+4112  &  1.734  &  0.68    &  7.737 & 28.9 & 0.28  &  9.4   &  0.33  &  $0.39$ $ $  $ $(MgII)\\
HE 1104$-$1805  &  2.32   &  0.73    &  8.244 & 21.7 & 2.23  &  14.5  &  0.67  &  $0.59$ $ $  $ $(H$_{\beta}$)  \\
Q 2237+0305     &  1.69   &  0.0395  &  3.660 & 8.11 & 0.23  &  13.6  &  1.68  &  $1.20$ $ $  $ $(H$_{\beta}$) 

\enddata

\tablecomments{Based on the source and lens redshifts $z_{s}$ and $z_{l}$, the Einstein radius $R_{E}$ can be computed. Here we report the estimates given by \cite{MosqueraKochanek2011} of $R_{E}$ as well as the Einstein radius and source crossing time scales $t_E$ and $t_s$, assuming a mean stellar mass in lens galaxies of $\langle M_{*} \rangle = 0.3$ $ M_{\odot}$, for comparison to the time span of the observations $\Delta t_{obs}$. The last column reports the estimated black hole mass and the emission lines used for the estimates by \cite{Morgan2010} (QJ 0158$-$4325, HE 0435$-$1223, SDSS 1004+4112), \cite{Peng2006} (SDSS 0924+0219) and \cite{Assef2011} (HE 1104$-$1805, Q 2237+0305).}

\end{deluxetable}


\begin{deluxetable}{lllllll}
\tablewidth{0pt}
\tabletypesize{\scriptsize}
\tablecolumns{7}


\centering

\tablecaption{RMS microlensing variability \label{tab:mean_rms_table}}


\tablehead{
\colhead{Object} &
\colhead{Pair} &
\colhead{N} &
\colhead{RMS, full} &
\colhead{RMS, hard} &
\colhead{RMS, soft} &
\colhead{p-value}
}

\startdata

QJ 0158$-$4325 & B/A & $12$ & $ 0.167 \pm 0.049 $ & $ 0.216 \pm 0.072 $ & $ 0.216 \pm 0.067 $ & $  1.00 $ \\
HE 1104$-$1805 & B/A & $15$ & $ 2.019 \pm 0.362 $ & $ 1.934 \pm 0.521 $ & $ 2.765 \pm 0.776 $ & $ <0.01 $ \\
HE 0435$-$1223 & B/A & $10$ & $ 0.216 \pm 0.049 $ & $ 0.234 \pm 0.073 $ & $ 0.215 \pm 0.069 $ & $  0.56 $ \\
HE 0435$-$1223 & C/A & $10$ & $ 0.224 \pm 0.043 $ & $ 0.276 \pm 0.080 $ & $ 0.169 \pm 0.050 $ & $ <0.01 $ \\
HE 0435$-$1223 & D/A & $10$ & $ 0.342 \pm 0.067 $ & $ 0.272 \pm 0.096 $ & $ 0.364 \pm 0.093 $ & $  0.04 $ \\
SDSS 0924+0219 & B/A & $ 6$ & $ 0.260 \pm 0.073 $ & $ 0.241 \pm 0.092 $ & $ 0.200 \pm 0.075 $ & $  0.42 $ \\
SDSS 0924+0219 & C/A & $ 6$ & $ 0.091 \pm 0.039 $ & $ 0.043 \pm 0.046 $ & $ 0.126 \pm 0.050 $ & $  0.01 $ \\
SDSS 0924+0219 & D/A & $ 6$ & $ 0.045 \pm 0.018 $ & $ 0.087 \pm 0.045 $ & $ 0.065 \pm 0.027 $ & $  0.34 $ \\
SDSS 1004+4112 & B/A & $11$ & $ 0.834 \pm 0.111 $ & $ 0.792 \pm 0.136 $ & $ 1.032 \pm 0.167 $ & $ <0.01 $ \\
SDSS 1004+4112 & C/A & $11$ & $ 1.914 \pm 0.524 $ & $ 1.766 \pm 0.642 $ & $ 2.149 \pm 0.733 $ & $  0.21 $ \\
SDSS 1004+4112 & D/A & $11$ & $ 3.272 \pm 0.802 $ & $ 3.022 \pm 1.036 $ & $ 3.821 \pm 1.110 $ & $  0.10 $ \\
Q 2237+0305    & B/A & $30$ & $ 0.114 \pm 0.025 $ & $ 0.147 \pm 0.031 $ & $ 0.108 \pm 0.027 $ & $ <0.01 $ \\
Q 2237+0305    & C/A & $30$ & $ 0.264 \pm 0.024 $ & $ 0.308 \pm 0.037 $ & $ 0.265 \pm 0.031 $ & $ <0.01 $ \\
Q 2237+0305    & D/A & $30$ & $ 0.182 \pm 0.014 $ & $ 0.196 \pm 0.021 $ & $ 0.224 \pm 0.035 $ & $ <0.01 $ \\

\enddata

\end{deluxetable}


\begin{deluxetable}{ccccccccc}
\tabletypesize{\scriptsize}
\setlength{\tabcolsep}{0.03in} 
\tablecolumns{13}
\tablewidth{0pt}
\tablecaption{Absorption-corrected count rates for QJ 0158$-$4325. \label{tab:0158count}}
\tablehead{
\colhead{ObsId}&
\colhead{Date} & 
\colhead{Exp} & 
\colhead{ $\rm{A_{full} }$ }& 
\colhead{$\rm{A_{soft}}$}&
\colhead{$\rm{A_{hard}}$}&
\colhead{ $\rm{B_{full} }$}& 
\colhead{$\rm{B_{soft}}$}&
\colhead{$\rm{B_{hard}}$}
}
\startdata

 11556 & 2009-Nov-06 & $ 5.03 $ & $   23.7 ^{+   3.2}_{  -6.6}$  & $   13.9 ^{+   2.6}_{  -2.5}$  & $    9.7 ^{+   1.8}_{  -1.9}$  & $   10.2 ^{+   4.2}_{  -2.6}$  & $    9.3 ^{+   2.4}_{  -2.2}$  & $    2.2 ^{+   0.8}_{  -0.7}$  \\
 11557 & 2010-Jan-12 & $ 5.02 $ & $   27.3 ^{+   4.1}_{  -2.8}$  & $   19.8 ^{+   3.0}_{  -3.3}$  & $    8.9 ^{+   1.7}_{  -1.6}$  & $   13.6 ^{+   2.5}_{  -2.1}$  & $   10.9 ^{+   1.9}_{  -2.2}$  & $    4.0 ^{+   1.2}_{  -0.9}$  \\
 11558 & 2010-Mar-10 & $ 5.04 $ & $   29.4 ^{+   3.9}_{  -4.4}$  & $   20.5 ^{+   5.4}_{  -2.9}$  & $    8.0 ^{+   2.1}_{  -1.8}$  & $   11.0 ^{+   2.3}_{  -2.0}$  & $    7.4 ^{+   1.7}_{  -1.3}$  & $    4.2 ^{+   1.3}_{  -0.9}$  \\
 11559 & 2010-May-23 & $ 4.94 $ & $   32.6 ^{+   2.9}_{  -3.9}$  & $   21.7 ^{+   3.0}_{  -2.9}$  & $   10.5 ^{+   1.8}_{  -2.2}$  & $    8.5 ^{+   1.6}_{  -1.5}$  & $    6.1 ^{+   1.5}_{  -1.3}$  & $    3.2 ^{+   0.9}_{  -1.0}$  \\
 11560 & 2010-Jul-28 & $ 4.95 $ & $   32.6 ^{+   2.9}_{  -3.7}$  & $   20.7 ^{+   2.6}_{  -3.3}$  & $   10.1 ^{+   2.4}_{  -1.6}$  & $    9.3 ^{+   1.7}_{  -1.7}$  & $    7.2 ^{+   1.6}_{  -1.6}$  & $    2.6 ^{+   1.0}_{  -0.8}$  \\
 11561 & 2010-Oct-06 & $ 4.95 $ & $   26.1 ^{+   3.0}_{  -3.4}$  & $   20.1 ^{+   2.4}_{  -3.1}$  & $    7.3 ^{+   1.9}_{  -1.8}$  & $   12.1 ^{+   1.8}_{  -1.8}$  & $   10.2 ^{+   1.9}_{  -2.3}$  & $    2.7 ^{+   1.0}_{  -0.8}$  \\
 14483 & 2013-Mar-26 & $ 18.6 $ & $   23.5 ^{+   1.4}_{  -4.2}$  & $   15.0 ^{+   1.1}_{  -5.0}$  & $    8.9 ^{+   0.8}_{  -1.2}$  & $    6.5 ^{+   0.7}_{  -0.7}$  & $    5.3 ^{+   0.7}_{  -0.7}$  & $    1.8 ^{+   0.4}_{  -0.3}$  \\
 14484 & 2013-Apr-24 & $ 18.6 $ & $   19.9 ^{+   1.5}_{  -2.1}$  & $   14.1 ^{+   1.0}_{  -1.1}$  & $    6.9 ^{+   0.7}_{  -0.7}$  & $    6.7 ^{+   0.7}_{  -1.4}$  & $    4.7 ^{+   0.7}_{  -0.7}$  & $    2.4 ^{+   0.4}_{  -0.4}$  \\
 14485 & 2013-Dec-05 & $ 18.6 $ & $   26.9 ^{+   1.7}_{  -5.0}$  & $   18.1 ^{+   2.4}_{  -2.3}$  & $    9.4 ^{+   0.9}_{  -1.3}$  & $    6.6 ^{+   2.2}_{  -0.7}$  & $    4.6 ^{+   0.8}_{  -0.8}$  & $    2.1 ^{+   0.4}_{  -0.4}$  \\
 14486 & 2013-Dec-29 & $ 18.6 $ & $   22.8 ^{+   2.3}_{  -2.1}$  & $   13.9 ^{+   1.1}_{  -1.2}$  & $    9.2 ^{+   1.2}_{  -1.4}$  & $    9.7 ^{+   1.0}_{  -0.9}$  & $    6.8 ^{+   0.8}_{  -0.7}$  & $    3.5 ^{+   0.6}_{  -0.5}$  \\
 14487 & 2014-May-30 & $ 18.6 $ & $   37.2 ^{+   1.5}_{  -1.6}$  & $   23.1 ^{+   1.5}_{  -2.0}$  & $   13.7 ^{+   1.1}_{  -1.5}$  & $    9.9 ^{+   0.9}_{  -0.9}$  & $    6.9 ^{+   0.8}_{  -0.7}$  & $    3.3 ^{+   0.6}_{  -0.6}$  \\
 14488 & 2014-Jun-11 & $ 18.6 $ & $   30.8 ^{+   2.1}_{  -4.7}$  & $   21.6 ^{+   2.9}_{  -5.9}$  & $   10.2 ^{+   1.6}_{  -1.2}$  & $   10.4 ^{+   2.6}_{  -1.6}$  & $    7.2 ^{+   2.6}_{  -2.6}$  & $    4.4 ^{+   0.7}_{  -0.7}$
\enddata

\tablecomments{Count rates are in units of $10^{-3}\,{\rm s}^{-1}$. \textit{Exp} reports the values stored in the header keyword EXPOSURE in units of $10^{3}\,{\rm s}$}

\end{deluxetable}

\begin{deluxetable}{ccccccccccccccc}
\tabletypesize{\scriptsize}
\setlength{\tabcolsep}{0.03in} 
\tablecolumns{13}
\tablewidth{0pt}
\tablecaption{Absorption-corrected count rates for HE 0435$-$1223.   \label{tab:0435count}}
\tablehead{
\colhead{ObsId}&
\colhead{Date} & 
\colhead{Exp\tablenotemark{b}} & 
\colhead{ $\rm{A_{full} }$ }& 
\colhead{$\rm{A_{soft}}$}&
\colhead{$\rm{A_{hard}}$}&
\colhead{ $\rm{B_{full} }$}& 
\colhead{$\rm{B_{soft}}$}&
\colhead{$\rm{B_{hard}}$}&
\colhead{ $\rm{C_{full} }$}& 
\colhead{$\rm{C_{soft}}$}&
\colhead{$\rm{C_{hard}}$}&
\colhead{ $\rm{D_{full} }$ }& 
\colhead{$\rm{D_{soft}}$}&
\colhead{$\rm{D_{hard}}$}
}
\startdata

  7761 & 2006-Dec-17 & $ 10.0 $ & $   28.1 ^{+   2.7}_{  -3.4}$  & $   17.8 ^{+   2.1}_{  -1.9}$  & $   10.0 ^{+   3.4}_{  -1.2}$  & $    9.5 ^{+   1.8}_{  -1.2}$  & $    7.5 ^{+   1.2}_{  -1.2}$  & $    3.2 ^{+   0.7}_{  -0.7}$  & $    9.0 ^{+   1.7}_{  -1.1}$  & $    6.9 ^{+   1.1}_{  -1.1}$  & $    3.8 ^{+   0.7}_{  -0.7}$  & $    8.6 ^{+   1.7}_{  -1.0}$  & $    5.4 ^{+   0.9}_{  -0.8}$  & $    4.6 ^{+   0.8}_{  -1.4}$  \\
 11550 & 2009-Dec-07 & $ 12.9 $ & $    8.4 ^{+   1.1}_{  -1.0}$  & $    5.3 ^{+   1.0}_{  -1.1}$  & $    3.9 ^{+   0.6}_{  -0.6}$  & $    9.7 ^{+   1.2}_{  -1.4}$  & $    6.2 ^{+   1.1}_{  -1.2}$  & $    3.7 ^{+   0.7}_{  -0.5}$  & $   10.1 ^{+   1.2}_{  -1.2}$  & $    5.4 ^{+   1.7}_{  -0.8}$  & $    4.1 ^{+   0.8}_{  -0.6}$  & $    8.6 ^{+   1.0}_{  -1.1}$  & $    5.4 ^{+   1.0}_{  -0.9}$  & $    2.9 ^{+   0.6}_{  -0.5}$  \\
 11551 & 2010-Jul-05 & $ 12.8 $ & $    7.8 ^{+   1.1}_{  -0.9}$  & $    4.6 ^{+   0.8}_{  -1.0}$  & $    3.8 ^{+   0.6}_{  -0.7}$  & $    7.0 ^{+   0.9}_{  -0.9}$  & $    4.1 ^{+   0.8}_{  -1.0}$  & $    3.1 ^{+   0.8}_{  -0.5}$  & $    8.2 ^{+   1.1}_{  -1.0}$  & $    3.6 ^{+   1.0}_{  -0.7}$  & $    4.5 ^{+   0.9}_{  -0.7}$  & $    7.4 ^{+   1.0}_{  -0.9}$  & $    4.5 ^{+   0.9}_{  -0.8}$  & $    3.4 ^{+   0.8}_{  -0.6}$  \\
 11552 & 2010-Oct-30 & $ 12.8 $ & $    5.2 ^{+   1.0}_{  -0.8}$  & $    3.0 ^{+   0.7}_{  -0.7}$  & $    2.4 ^{+   0.6}_{  -0.5}$  & $    4.5 ^{+   0.9}_{  -0.8}$  & $    1.8 ^{+   0.5}_{  -0.5}$  & $    2.8 ^{+   0.6}_{  -0.5}$  & $    5.6 ^{+   0.9}_{  -0.8}$  & $    2.1 ^{+   0.7}_{  -0.5}$  & $    3.4 ^{+   0.7}_{  -0.6}$  & $    4.8 ^{+   0.8}_{  -0.7}$  & $    2.4 ^{+   0.8}_{  -0.6}$  & $    2.4 ^{+   0.5}_{  -0.4}$  \\
 14489 & 2012-Nov-28 & $ 37.2 $ & $   11.7 ^{+   0.6}_{  -0.8}$  & $    7.5 ^{+   1.2}_{  -1.2}$  & $    4.7 ^{+   0.9}_{  -0.5}$  & $    7.2 ^{+   0.5}_{  -0.6}$  & $    4.5 ^{+   0.5}_{  -0.9}$  & $    2.8 ^{+   0.4}_{  -0.4}$  & $   10.0 ^{+   0.6}_{  -0.6}$  & $    5.7 ^{+   1.6}_{  -0.7}$  & $    4.7 ^{+   0.8}_{  -0.7}$  & $    8.5 ^{+   0.7}_{  -0.6}$  & $    4.1 ^{+   1.3}_{  -0.5}$  & $    4.2 ^{+   0.4}_{  -0.5}$  \\
 14490 & 2013-Apr-01 & $ 36.2 $ & $   11.6 ^{+   0.6}_{  -2.3}$  & $    8.1 ^{+   0.9}_{  -0.8}$  & $    4.6 ^{+   0.4}_{  -0.4}$  & $    7.3 ^{+   0.5}_{  -2.1}$  & $    4.4 ^{+   0.6}_{  -0.5}$  & $    2.9 ^{+   0.3}_{  -0.3}$  & $   10.7 ^{+   2.8}_{  -2.1}$  & $    6.9 ^{+   0.7}_{  -0.7}$  & $    4.5 ^{+   0.4}_{  -0.4}$  & $    6.3 ^{+   0.5}_{  -1.5}$  & $    4.3 ^{+   0.5}_{  -0.5}$  & $    2.7 ^{+   0.3}_{  -0.3}$  \\
 14491 & 2013-Aug-14 & $ 36.2 $ & $   11.9 ^{+   0.7}_{  -0.8}$  & $    7.8 ^{+   0.6}_{  -0.8}$  & $    6.0 ^{+   0.4}_{  -0.6}$  & $    6.7 ^{+   0.5}_{  -0.6}$  & $    3.5 ^{+   0.5}_{  -0.4}$  & $    2.9 ^{+   0.4}_{  -0.3}$  & $    9.5 ^{+   0.8}_{  -0.9}$  & $    5.1 ^{+   0.6}_{  -0.5}$  & $    4.0 ^{+   0.5}_{  -0.4}$  & $    9.0 ^{+   0.6}_{  -2.5}$  & $    4.9 ^{+   0.6}_{  -0.6}$  & $    3.8 ^{+   0.5}_{  -0.4}$  \\
 14492 & 2013-Sep-22 & $ 35.5 $ & $   10.6 ^{+   0.8}_{  -0.9}$  & $    5.8 ^{+   0.7}_{  -0.6}$  & $    4.8 ^{+   1.3}_{  -1.1}$  & $    5.6 ^{+   1.8}_{  -0.5}$  & $    3.6 ^{+   0.5}_{  -0.5}$  & $    2.5 ^{+   1.1}_{  -0.3}$  & $    8.2 ^{+   0.8}_{  -0.6}$  & $    5.4 ^{+   0.5}_{  -0.9}$  & $    3.6 ^{+   0.4}_{  -0.8}$  & $    6.3 ^{+   0.4}_{  -0.4}$  & $    4.0 ^{+   0.5}_{  -0.5}$  & $    2.9 ^{+   0.4}_{  -0.6}$  \\
 14493 & 2014-Mar-08 & $ 36.2 $ & $   11.6 ^{+   1.0}_{  -0.8}$  & $    6.5 ^{+   1.2}_{  -1.0}$  & $    5.7 ^{+   0.7}_{  -0.6}$  & $    6.3 ^{+   0.5}_{  -0.5}$  & $    3.4 ^{+   0.5}_{  -0.7}$  & $    3.0 ^{+   0.4}_{  -0.4}$  & $   11.1 ^{+   0.8}_{  -1.0}$  & $    5.7 ^{+   0.6}_{  -0.8}$  & $    5.2 ^{+   0.5}_{  -0.6}$  & $    6.1 ^{+   0.5}_{  -0.5}$  & $    2.8 ^{+   0.4}_{  -0.4}$  & $    3.3 ^{+   0.3}_{  -0.3}$  \\
 14494 & 2014-Apr-10 & $ 36.2 $ & $    9.9 ^{+   0.9}_{  -0.8}$  & $    5.4 ^{+   0.6}_{  -0.5}$  & $    4.8 ^{+   0.4}_{  -0.4}$  & $    6.8 ^{+   0.7}_{  -0.7}$  & $    4.3 ^{+   0.5}_{  -0.5}$  & $    3.4 ^{+   0.3}_{  -0.3}$  & $    8.4 ^{+   0.7}_{  -0.7}$  & $    4.9 ^{+   0.5}_{  -0.5}$  & $    3.9 ^{+   0.4}_{  -0.4}$  & $    7.2 ^{+   0.6}_{  -0.7}$  & $    4.6 ^{+   0.4}_{  -0.4}$  & $    3.1 ^{+   0.4}_{  -0.4}$  \\
\enddata

\tablecomments{Count rates are in units of $10^{-3}\,{\rm s}^{-1}$. \textit{Exp} reports the values stored in the header keyword EXPOSURE in units of $10^{3}\,{\rm s}$}

\end{deluxetable}

\begin{deluxetable}{ccccccccccccccc}
\tabletypesize{\scriptsize}
\setlength{\tabcolsep}{0.03in} 
\tablecolumns{13}
\tablewidth{0pt}
\tablecaption{Absorption-corrected count rates for SDSS 0924+0219.   \label{tab:0924count}}
\tablehead{
\colhead{ObsId}&
\colhead{Date} & 
\colhead{Exp\tablenotemark{b}} & 
\colhead{ $\rm{A_{full} }$ }& 
\colhead{$\rm{A_{soft}}$}&
\colhead{$\rm{A_{hard}}$}&
\colhead{ $\rm{B_{full} }$}& 
\colhead{$\rm{B_{soft}}$}&
\colhead{$\rm{B_{hard}}$}&
\colhead{ $\rm{C_{full} }$}& 
\colhead{$\rm{C_{soft}}$}&
\colhead{$\rm{C_{hard}}$}&
\colhead{ $\rm{D_{full} }$ }& 
\colhead{$\rm{D_{soft}}$}&
\colhead{$\rm{D_{hard}}$}
}
\startdata

  5604 & 2005-Feb-24 & $ 18.0 $ & $    4.6 ^{+   1.2}_{  -0.6}$  & $    3.5 ^{+   0.8}_{  -0.7}$  & $    1.5 ^{+   0.6}_{  -0.3}$  & $    1.7 ^{+   0.4}_{  -0.4}$  & $    1.3 ^{+   0.4}_{  -0.3}$  & $    0.4 ^{+   0.2}_{  -0.2}$  & $    0.7 ^{+   0.2}_{  -0.2}$  & $    0.5 ^{+   0.2}_{  -0.2}$  & $    0.1 ^{+   0.1}_{  -0.1}$  & $    0.6 ^{+   0.3}_{  -0.2}$  & $    0.6 ^{+   0.3}_{  -0.2}$  & $    0.1 ^{+   0.1}_{  -0.1}$  \\
 11562 & 2010-Jan-13 & $ 21.5 $ & $   10.1 ^{+   1.3}_{  -1.6}$  & $    8.2 ^{+   0.8}_{  -1.8}$  & $    2.7 ^{+   0.5}_{  -0.4}$  & $    2.5 ^{+   0.6}_{  -0.5}$  & $    1.7 ^{+   0.3}_{  -0.3}$  & $    0.8 ^{+   0.3}_{  -0.2}$  & $    0.6 ^{+   0.2}_{  -0.2}$  & $    0.6 ^{+   0.2}_{  -0.2}$  & $    0.1 ^{+   0.1}_{  -0.1}$  & $    1.7 ^{+   0.4}_{  -0.4}$  & $    1.5 ^{+   0.4}_{  -0.3}$  & $    0.4 ^{+   0.2}_{  -0.2}$  \\
 11563 & 2010-Mar-11 & $ 21.3 $ & $    6.8 ^{+   0.7}_{  -1.1}$  & $    4.2 ^{+   0.8}_{  -0.6}$  & $    2.2 ^{+   0.5}_{  -0.5}$  & $    0.7 ^{+   0.2}_{  -0.2}$  & $    0.5 ^{+   0.2}_{  -0.2}$  & $    0.3 ^{+   0.2}_{  -0.1}$  & $    0.8 ^{+   0.3}_{  -0.2}$  & $    0.8 ^{+   0.3}_{  -0.2}$  & $    0.1 ^{+   0.1}_{  -0.1}$  & $    0.7 ^{+   0.3}_{  -0.2}$  & $    0.9 ^{+   0.3}_{  -0.3}$  & $    0.1 ^{+   0.1}_{  -0.1}$  \\
 11564 & 2010-May-10 & $ 21.6 $ & $   16.0 ^{+   1.0}_{  -1.0}$  & $   10.7 ^{+   1.2}_{  -1.4}$  & $    5.0 ^{+   0.7}_{  -0.8}$  & $    2.3 ^{+   0.4}_{  -0.3}$  & $    1.9 ^{+   0.4}_{  -0.3}$  & $    0.7 ^{+   0.2}_{  -0.2}$  & $    0.8 ^{+   0.2}_{  -0.2}$  & $    0.4 ^{+   0.2}_{  -0.1}$  & $    0.4 ^{+   0.2}_{  -0.1}$  & $    1.1 ^{+   0.3}_{  -0.3}$  & $    0.7 ^{+   0.3}_{  -0.2}$  & $    0.3 ^{+   0.2}_{  -0.1}$  \\
 11565 & 2010-Jun-23 & $ 21.6 $ & $   10.2 ^{+   2.1}_{  -1.0}$  & $    7.7 ^{+   1.1}_{  -1.0}$  & $    3.7 ^{+   0.5}_{  -0.6}$  & $    4.1 ^{+   0.6}_{  -0.9}$  & $    2.5 ^{+   0.5}_{  -0.5}$  & $    1.5 ^{+   0.4}_{  -0.3}$  & $    0.7 ^{+   0.3}_{  -0.2}$  & $    0.4 ^{+   0.2}_{  -0.1}$  & $    0.2 ^{+   0.1}_{  -0.1}$  & $    0.9 ^{+   0.5}_{  -0.3}$  & $    0.8 ^{+   0.3}_{  -0.2}$  & $    0.2 ^{+   0.2}_{  -0.1}$  \\
 11566 & 2010-Oct-06 & $ 21.5 $ & $    7.1 ^{+   1.7}_{  -0.7}$  & $    5.9 ^{+   0.7}_{  -1.0}$  & $    2.4 ^{+   0.4}_{  -0.8}$  & $    2.0 ^{+   0.5}_{  -0.4}$  & $    1.6 ^{+   0.5}_{  -0.3}$  & $    0.4 ^{+   0.4}_{  -0.1}$  & $    0.7 ^{+   0.3}_{  -0.3}$  & $    0.5 ^{+   0.2}_{  -0.2}$  & $    0.2 ^{+   0.3}_{  -0.1}$  & $    1.1 ^{+   0.4}_{  -0.3}$  & $    0.5 ^{+   0.3}_{  -0.2}$  & $    0.6 ^{+   0.4}_{  -0.2}$  \\
\enddata

\tablecomments{Count rates are in units of $10^{-3}\,{\rm s}^{-1}$. \textit{Exp} reports the values stored in the header keyword EXPOSURE in units of $10^{3}\,{\rm s}$}

\end{deluxetable}

\begin{deluxetable}{ccccccccccccccc}
\tabletypesize{\scriptsize}
\setlength{\tabcolsep}{0.03in} 
\tablecolumns{13}
\tablewidth{0pt}
\tablecaption{Absorption-corrected count rates for SDSS 1004+4112.   \label{tab:1004count}}
\tablehead{
\colhead{ObsId}&
\colhead{Date} & 
\colhead{Exp\tablenotemark{b}} & 
\colhead{ $\rm{A_{full} }$ }& 
\colhead{$\rm{A_{soft}}$}&
\colhead{$\rm{A_{hard}}$}&
\colhead{ $\rm{B_{full} }$}& 
\colhead{$\rm{B_{soft}}$}&
\colhead{$\rm{B_{hard}}$}&
\colhead{ $\rm{C_{full} }$}& 
\colhead{$\rm{C_{soft}}$}&
\colhead{$\rm{C_{hard}}$}&
\colhead{ $\rm{D_{full} }$ }& 
\colhead{$\rm{D_{soft}}$}&
\colhead{$\rm{D_{hard}}$}
}
\startdata

 11556 & 2005-Jan-02 & $ 80.1 $ & $   16.9{\pm    0.5}$  & $   10.5{\pm    0.4}$  & $    7.0{\pm    0.3}$  & $   21.6{\pm    0.5}$  & $   13.6{\pm    0.4}$  & $    8.4{\pm    0.3}$  & $   17.6{\pm    0.5}$  & $   10.6{\pm    0.4}$  & $    7.2{\pm    0.3}$  & $    9.9{\pm    0.4}$  & $    5.8{\pm    0.3}$  & $    4.3{\pm    0.2}$ \\
 11557 & 2010-Mar-08 & $  5.96 $ & $   11.4{\pm    1.4}$  & $    5.5{\pm    1.0}$  & $    6.1{\pm    1.0}$  & $   16.0{\pm    1.7}$  & $   10.8{\pm    1.4}$  & $    5.5{\pm    1.0}$  & $   12.0{\pm    1.4}$  & $    6.2{\pm    1.1}$  & $    5.8{\pm    1.0}$  & $   15.3{\pm    1.6}$  & $   10.5{\pm    1.4}$  & $    5.1{\pm    0.9}$ \\
 11558 & 2010-Jun-19 & $  5.96 $ & $    8.8{\pm    1.3}$  & $    5.5{\pm    1.0}$  & $    3.6{\pm    0.8}$  & $    8.6{\pm    1.2}$  & $    5.1{\pm    1.0}$  & $    3.6{\pm    0.8}$  & $   17.0{\pm    1.7}$  & $   10.2{\pm    1.4}$  & $    7.1{\pm    1.1}$  & $   11.0{\pm    1.4}$  & $    6.2{\pm    1.1}$  & $    4.9{\pm    0.9}$ \\
 11559 & 2010-Sep-23 & $  5.96 $ & $    9.2{\pm    1.3}$  & $    4.7{\pm    1.0}$  & $    4.6{\pm    0.9}$  & $   12.0{\pm    1.4}$  & $    5.9{\pm    1.0}$  & $    6.3{\pm    1.0}$  & $   11.4{\pm    1.4}$  & $    6.2{\pm    1.1}$  & $    5.3{\pm    1.0}$  & $    8.3{\pm    1.2}$  & $    4.1{\pm    0.9}$  & $    4.3{\pm    0.9}$ \\
 11560 & 2011-Jan-30 & $  5.96 $ & $    5.3{\pm    1.0}$  & $    2.8{\pm    0.7}$  & $    2.6{\pm    0.7}$  & $    7.9{\pm    1.2}$  & $    4.0{\pm    0.9}$  & $    4.0{\pm    0.8}$  & $   19.6{\pm    1.8}$  & $   10.7{\pm    1.4}$  & $    9.2{\pm    1.2}$  & $   14.4{\pm    1.6}$  & $    7.8{\pm    1.2}$  & $    6.8{\pm    1.1}$ \\
 11561 & 2013-Jan-28 & $ 24.7 $ & $    9.0{\pm    0.6}$  & $    4.1{\pm    0.4}$  & $    5.0{\pm    0.5}$  & $   19.4{\pm    0.9}$  & $   10.5{\pm    0.7}$  & $    9.1{\pm    0.6}$  & $   14.6{\pm    0.8}$  & $    7.4{\pm    0.6}$  & $    7.4{\pm    0.5}$  & $    8.2{\pm    0.6}$  & $    4.3{\pm    0.4}$  & $    4.0{\pm    0.4}$ \\
 14483 & 2013-Mar-01 & $ 24.7 $ & $    9.4{\pm    0.6}$  & $    4.7{\pm    0.5}$  & $    4.9{\pm    0.5}$  & $   19.9{\pm    0.9}$  & $   10.5{\pm    0.7}$  & $    9.6{\pm    0.6}$  & $   15.2{\pm    0.8}$  & $    7.5{\pm    0.6}$  & $    7.9{\pm    0.6}$  & $    7.0{\pm    0.5}$  & $    3.4{\pm    0.4}$  & $    3.6{\pm    0.4}$ \\
 14484 & 2013-Oct-05 & $ 24.1 $ & $   10.2{\pm    0.7}$  & $    4.9{\pm    0.5}$  & $    5.4{\pm    0.5}$  & $   16.9{\pm    0.9}$  & $    9.1{\pm    0.6}$  & $    8.0{\pm    0.6}$  & $   15.4{\pm    0.8}$  & $    7.4{\pm    0.6}$  & $    8.1{\pm    0.6}$  & $    6.0{\pm    0.5}$  & $    3.1{\pm    0.4}$  & $    3.0{\pm    0.4}$ \\
 14485 & 2013-Nov-16 & $ 23.8 $ & $    8.9{\pm    0.6}$  & $    4.0{\pm    0.4}$  & $    5.0{\pm    0.5}$  & $   17.4{\pm    0.9}$  & $    9.4{\pm    0.7}$  & $    8.2{\pm    0.6}$  & $   17.3{\pm    0.9}$  & $    9.7{\pm    0.7}$  & $    7.8{\pm    0.6}$  & $   10.3{\pm    0.7}$  & $    4.7{\pm    0.5}$  & $    5.6{\pm    0.5}$ \\
 14486 & 2014-Apr-30 & $ 23.3 $ & $    8.0{\pm    0.6}$  & $    4.1{\pm    0.5}$  & $    4.1{\pm    0.4}$  & $   14.1{\pm    0.8}$  & $    6.9{\pm    0.6}$  & $    7.3{\pm    0.6}$  & $   10.8{\pm    0.7}$  & $    5.7{\pm    0.5}$  & $    5.2{\pm    0.5}$  & $    5.3{\pm    0.5}$  & $    2.3{\pm    0.3}$  & $    3.0{\pm    0.4}$ \\
 14487 & 2014-Jun-02 & $ 24.7 $ & $    7.1{\pm    0.6}$  & $    3.2{\pm    0.4}$  & $    4.0{\pm    0.4}$  & $   20.8{\pm    0.9}$  & $    9.9{\pm    0.7}$  & $   11.1{\pm    0.7}$  & $   11.5{\pm    0.7}$  & $    5.2{\pm    0.5}$  & $    6.3{\pm    0.5}$  & $    4.5{\pm    0.4}$  & $    2.1{\pm    0.3}$  & $    2.4{\pm    0.3}$ \\
\enddata

\tablecomments{Count rates are in units of $10^{-3}\,{\rm s}^{-1}$. \textit{Exp} reports the values stored in the header keyword EXPOSURE in units of $10^{3}\,{\rm s}$}

\end{deluxetable}

\begin{deluxetable}{ccccccccc}
\tabletypesize{\scriptsize}
\setlength{\tabcolsep}{0.03in} 
\tablecolumns{13}
\tablewidth{0pt}
\tablecaption{Absorption-corrected count rates for HE 1104$-$1805.   \label{tab:1104count}}
\tablehead{
\colhead{ObsId}&
\colhead{Date} & 
\colhead{Exp\tablenotemark{b}} & 
\colhead{ $\rm{A_{full} }$ }& 
\colhead{$\rm{A_{soft}}$}&
\colhead{$\rm{A_{hard}}$}&
\colhead{ $\rm{B_{full} }$}& 
\colhead{$\rm{B_{soft}}$}&
\colhead{$\rm{B_{hard}}$}
}
\startdata

   375 & 2000-Jun-11 & $ 47.4 $ & $   21.5 ^{+   0.9}_{  -1.0}$  & $   13.6 ^{+   0.7}_{  -0.7}$  & $   10.2 ^{+   0.6}_{  -0.7}$  & $   12.0 ^{+   0.5}_{  -0.5}$  & $    8.2 ^{+   0.6}_{  -0.6}$  & $    4.6 ^{+   0.4}_{  -0.4}$  \\
  6917 & 2006-Mar-15 & $ 4.55 $ & $    8.0 ^{+   2.5}_{  -1.6}$  & $    3.6 ^{+   1.2}_{  -1.3}$  & $    6.0 ^{+   1.3}_{  -1.2}$  & $   12.3 ^{+   2.3}_{  -2.5}$  & $    8.8 ^{+   1.8}_{  -2.5}$  & $    6.2 ^{+   1.3}_{  -1.2}$  \\
  6918 & 2006-Feb-16 & $ 4.96 $ & $    7.8 ^{+   1.4}_{  -1.4}$  & $    2.9 ^{+   1.3}_{  -0.8}$  & $    4.2 ^{+   1.2}_{  -1.3}$  & $   17.5 ^{+   2.3}_{  -2.9}$  & $    6.1 ^{+   2.0}_{  -1.1}$  & $    9.8 ^{+   1.9}_{  -2.1}$  \\
  6919 & 2006-Apr-09 & $ 4.87 $ & $    7.2 ^{+   1.4}_{  -1.8}$  & $    2.8 ^{+   1.1}_{  -1.0}$  & $    4.4 ^{+   1.1}_{  -1.1}$  & $   12.0 ^{+   2.1}_{  -2.8}$  & $    8.0 ^{+   1.6}_{  -2.2}$  & $    5.7 ^{+   1.3}_{  -1.8}$  \\
  6920 & 2006-Oct-31 & $ 5.01 $ & $   12.6 ^{+   1.9}_{  -2.1}$  & $    5.6 ^{+   1.9}_{  -1.4}$  & $    5.8 ^{+   1.4}_{  -0.9}$  & $    7.4 ^{+   1.7}_{  -1.5}$  & $    4.7 ^{+   1.5}_{  -1.1}$  & $    3.0 ^{+   1.0}_{  -0.7}$  \\
  6921 & 2006-Nov-08 & $ 4.92 $ & $   10.9 ^{+   2.1}_{  -1.8}$  & $    5.2 ^{+   1.6}_{  -1.5}$  & $    6.0 ^{+   1.4}_{  -1.1}$  & $    9.3 ^{+   1.8}_{  -1.5}$  & $    5.3 ^{+   1.5}_{  -1.4}$  & $    4.8 ^{+   1.4}_{  -1.0}$  \\
 11553 & 2010-Feb-09 & $ 12.8 $ & $    8.7 ^{+   1.0}_{  -1.2}$  & $    3.8 ^{+   1.0}_{  -0.7}$  & $    4.3 ^{+   1.1}_{  -0.7}$  & $   12.6 ^{+   1.1}_{  -1.2}$  & $    5.7 ^{+   1.3}_{  -1.1}$  & $    6.5 ^{+   1.1}_{  -1.2}$  \\
 11554 & 2010-Jul-12 & $ 12.8 $ & $   19.7 ^{+   1.9}_{  -3.8}$  & $    8.8 ^{+   1.4}_{  -1.9}$  & $   12.1 ^{+   1.2}_{  -1.6}$  & $    7.8 ^{+   1.3}_{  -1.7}$  & $    3.8 ^{+   0.7}_{  -1.3}$  & $    4.5 ^{+   0.7}_{  -0.7}$  \\
 11555 & 2010-Dec-20 & $ 12.8 $ & $   10.2 ^{+   1.0}_{  -1.2}$  & $    4.0 ^{+   1.1}_{  -0.8}$  & $    5.7 ^{+   1.0}_{  -0.9}$  & $   11.7 ^{+   1.3}_{  -1.3}$  & $    5.1 ^{+   1.3}_{  -1.1}$  & $    6.3 ^{+   1.1}_{  -0.9}$  \\
 14501 & 2013-Mar-01 & $ 13.7 $ & $    8.1 ^{+   1.0}_{  -1.0}$  & $    2.9 ^{+   0.6}_{  -0.5}$  & $    5.3 ^{+   0.7}_{  -0.7}$  & $    7.1 ^{+   0.9}_{  -0.8}$  & $    2.7 ^{+   0.7}_{  -0.5}$  & $    4.8 ^{+   0.7}_{  -0.7}$  \\
 14502 & 2013-Jul-26 & $ 13.7 $ & $    7.1 ^{+   0.8}_{  -1.4}$  & $    2.3 ^{+   0.6}_{  -0.4}$  & $    4.6 ^{+   0.6}_{  -1.2}$  & $   11.3 ^{+   1.2}_{  -1.7}$  & $    4.5 ^{+   0.9}_{  -0.7}$  & $    6.6 ^{+   0.9}_{  -1.3}$  \\
 14503 & 2013-Dec-08 & $ 12.8 $ & $    6.3 ^{+   1.0}_{  -0.8}$  & $    2.6 ^{+   0.9}_{  -0.5}$  & $    3.7 ^{+   0.7}_{  -0.5}$  & $    5.8 ^{+   0.8}_{  -0.7}$  & $    2.8 ^{+   0.8}_{  -0.7}$  & $    3.3 ^{+   0.6}_{  -0.5}$  \\
 14504 & 2014-Mar-25 & $ 13.7 $ & $   10.6 ^{+   1.2}_{  -1.2}$  & $    3.8 ^{+   0.7}_{  -2.4}$  & $    6.7 ^{+   0.9}_{  -0.8}$  & $    8.7 ^{+   0.9}_{  -1.1}$  & $    3.2 ^{+   0.7}_{  -1.2}$  & $    5.3 ^{+   0.8}_{  -0.8}$  \\
 14505 & 2014-Jul-14 & $ 13.7 $ & $   13.0 ^{+   1.5}_{  -1.9}$  & $    5.0 ^{+   0.7}_{  -1.3}$  & $    9.2 ^{+   0.9}_{  -1.5}$  & $    6.2 ^{+   1.7}_{  -0.8}$  & $    2.7 ^{+   0.6}_{  -0.7}$  & $    4.3 ^{+   0.6}_{  -0.6}$  \\
 14506 & 2014-Dec-02 & $ 13.7 $ & $    4.7 ^{+   0.7}_{  -0.8}$  & $    1.5 ^{+   0.5}_{  -0.4}$  & $    3.0 ^{+   0.6}_{  -0.4}$  & $    7.1 ^{+   1.1}_{  -1.4}$  & $    2.8 ^{+   0.7}_{  -0.5}$  & $    3.7 ^{+   0.8}_{  -0.5}$  \\
\enddata

\tablecomments{Count rates are in units of $10^{-3}\,{\rm s}^{-1}$. \textit{Exp} reports the values stored in the header keyword EXPOSURE in units of $10^{3}\,{\rm s}$}

\end{deluxetable}

\begin{deluxetable}{ccccccccccccccc}
\tabletypesize{\scriptsize}
\setlength{\tabcolsep}{0.03in} 
\tablecolumns{13}
\tablewidth{0pt}
\tablecaption{Absorption-corrected count rates for Q~2237+0305.   \label{tab:2237count}}
\tablehead{
\colhead{ObsId}&
\colhead{Date} & 
\colhead{Exp\tablenotemark{b}} & 
\colhead{ $\rm{A_{full} }$ }& 
\colhead{$\rm{A_{soft}}$}&
\colhead{$\rm{A_{hard}}$}&
\colhead{ $\rm{B_{full} }$}& 
\colhead{$\rm{B_{soft}}$}&
\colhead{$\rm{B_{hard}}$}&
\colhead{ $\rm{C_{full} }$}& 
\colhead{$\rm{C_{soft}}$}&
\colhead{$\rm{C_{hard}}$}&
\colhead{ $\rm{D_{full} }$ }& 
\colhead{$\rm{D_{soft}}$}&
\colhead{$\rm{D_{hard}}$}
}
\startdata

   431 & 2000-Sep-07 & $ 30.3 $ & $   56.7 ^{+   2.1}_{  -2.3}$  & $   53.4 ^{+   3.8}_{  -4.2}$  & $   18.7 ^{+   2.3}_{  -2.0}$  & $    9.9 ^{+   0.7}_{  -0.7}$  & $    9.6 ^{+   1.0}_{  -1.0}$  & $    3.4 ^{+   0.4}_{  -0.6}$  & $   22.8 ^{+   1.2}_{  -1.1}$  & $   25.9 ^{+   1.5}_{  -1.6}$  & $    6.9 ^{+   2.0}_{  -0.6}$  & $    9.3 ^{+   0.7}_{  -0.7}$  & $    8.6 ^{+   1.8}_{  -0.9}$  & $    3.2 ^{+   0.9}_{  -0.4}$  \\
  1632 & 2001-Dec-08 & $ 9.54 $ & $   43.0 ^{+   3.5}_{  -6.3}$  & $   36.7 ^{+   4.3}_{  -4.0}$  & $   16.7 ^{+   1.6}_{  -2.3}$  & $    7.7 ^{+   2.0}_{  -1.0}$  & $    8.7 ^{+   1.5}_{  -1.5}$  & $    2.4 ^{+   0.6}_{  -0.5}$  & $   13.2 ^{+   1.8}_{  -1.5}$  & $   12.1 ^{+   2.3}_{  -2.1}$  & $    5.8 ^{+   1.0}_{  -0.9}$  & $    7.8 ^{+   1.2}_{  -1.1}$  & $    8.1 ^{+   1.5}_{  -1.4}$  & $    2.5 ^{+   0.6}_{  -0.5}$  \\
  6831 & 2006-Jan-10 & $ 7.27 $ & $   20.2 ^{+   2.0}_{  -5.5}$  & $   14.4 ^{+   2.7}_{  -2.0}$  & $    8.4 ^{+   1.3}_{  -1.6}$  & $   14.3 ^{+   1.8}_{  -3.1}$  & $    8.1 ^{+   1.7}_{  -3.7}$  & $    7.3 ^{+   1.3}_{  -1.5}$  & $    5.1 ^{+   1.1}_{  -1.0}$  & $    3.7 ^{+   1.3}_{  -1.1}$  & $    2.2 ^{+   0.7}_{  -0.5}$  & $    7.0 ^{+   3.2}_{  -1.2}$  & $    6.9 ^{+   1.6}_{  -4.4}$  & $    2.7 ^{+   0.9}_{  -0.6}$  \\
  6832 & 2006-May-01 & $ 7.94 $ & $   35.7 ^{+   6.9}_{  -3.7}$  & $   29.2 ^{+   5.0}_{  -5.3}$  & $   15.4 ^{+   2.3}_{  -2.1}$  & $   16.8 ^{+   2.3}_{  -1.8}$  & $   12.5 ^{+   2.3}_{  -1.8}$  & $    7.6 ^{+   1.3}_{  -1.2}$  & $   16.3 ^{+   2.1}_{  -3.4}$  & $   14.9 ^{+   2.9}_{  -2.8}$  & $    5.5 ^{+   0.9}_{  -0.8}$  & $   12.1 ^{+   1.7}_{  -3.9}$  & $    8.5 ^{+   2.0}_{  -1.7}$  & $    4.6 ^{+   1.0}_{  -0.8}$  \\
  6833 & 2006-May-27 & $ 7.95 $ & $   21.1 ^{+   3.0}_{  -2.9}$  & $   17.6 ^{+   2.9}_{  -3.8}$  & $    7.9 ^{+   1.7}_{  -1.2}$  & $    7.7 ^{+   1.5}_{  -1.2}$  & $    5.9 ^{+   1.7}_{  -1.5}$  & $    3.5 ^{+   1.1}_{  -0.7}$  & $    8.6 ^{+   1.5}_{  -1.6}$  & $    6.8 ^{+   1.8}_{  -2.1}$  & $    3.4 ^{+   1.2}_{  -0.6}$  & $    4.1 ^{+   1.1}_{  -0.9}$  & $    2.7 ^{+   2.2}_{  -0.8}$  & $    1.8 ^{+   0.6}_{  -0.5}$  \\
  6834 & 2006-Jun-25 & $ 7.94 $ & $   45.9 ^{+   4.0}_{  -4.2}$  & $   40.4 ^{+   5.3}_{  -5.0}$  & $   18.5 ^{+   2.8}_{  -3.1}$  & $   19.4 ^{+   2.0}_{  -2.1}$  & $   16.2 ^{+   2.3}_{  -2.4}$  & $    6.7 ^{+   2.0}_{  -1.1}$  & $   13.2 ^{+   1.8}_{  -1.7}$  & $   11.1 ^{+   2.2}_{  -2.1}$  & $    5.0 ^{+   1.7}_{  -0.8}$  & $   10.8 ^{+   1.5}_{  -1.3}$  & $    6.4 ^{+   1.5}_{  -1.3}$  & $    4.9 ^{+   1.1}_{  -0.8}$  \\
  6835 & 2006-Jul-21 & $ 7.87 $ & $   51.5 ^{+   5.4}_{  -3.2}$  & $   32.6 ^{+   4.1}_{  -4.7}$  & $   26.6 ^{+   3.2}_{  -2.6}$  & $   14.4 ^{+   1.6}_{  -2.7}$  & $    9.9 ^{+   1.7}_{  -1.5}$  & $    5.3 ^{+   0.9}_{  -0.9}$  & $   10.4 ^{+   1.6}_{  -1.7}$  & $    8.7 ^{+   1.9}_{  -1.7}$  & $    4.4 ^{+   1.0}_{  -1.0}$  & $    8.8 ^{+   1.4}_{  -1.5}$  & $    6.3 ^{+   1.6}_{  -1.3}$  & $    4.4 ^{+   0.9}_{  -0.9}$  \\
  6836 & 2006-Aug-17 & $ 7.93 $ & $   26.6 ^{+   3.3}_{  -2.2}$  & $   25.2 ^{+   3.4}_{  -7.6}$  & $   11.8 ^{+   1.9}_{  -1.3}$  & $    8.5 ^{+   1.6}_{  -1.2}$  & $    5.7 ^{+   1.4}_{  -1.9}$  & $    4.3 ^{+   0.9}_{  -0.7}$  & $    8.5 ^{+   1.5}_{  -1.4}$  & $    5.5 ^{+   1.6}_{  -1.5}$  & $    4.0 ^{+   0.9}_{  -0.7}$  & $    6.7 ^{+   1.2}_{  -1.1}$  & $    3.1 ^{+   2.7}_{  -0.8}$  & $    3.4 ^{+   0.8}_{  -0.7}$  \\
  6837 & 2006-Sep-16 & $ 7.95 $ & $   26.7 ^{+   6.5}_{  -2.6}$  & $   22.3 ^{+   4.0}_{  -3.7}$  & $   12.2 ^{+   1.6}_{  -1.9}$  & $   12.6 ^{+   2.0}_{  -1.9}$  & $    8.3 ^{+   1.9}_{  -1.5}$  & $    5.3 ^{+   1.0}_{  -0.8}$  & $    6.2 ^{+   1.5}_{  -1.1}$  & $    4.6 ^{+   1.4}_{  -1.0}$  & $    2.9 ^{+   0.8}_{  -0.6}$  & $    6.7 ^{+   1.3}_{  -1.0}$  & $    5.0 ^{+   1.3}_{  -1.1}$  & $    2.9 ^{+   0.8}_{  -0.6}$  \\
  6838 & 2006-Oct-09 & $ 7.99 $ & $   25.8 ^{+   2.9}_{  -3.3}$  & $   26.5 ^{+   3.2}_{  -4.4}$  & $    8.2 ^{+   1.5}_{  -1.3}$  & $    9.3 ^{+   1.7}_{  -1.2}$  & $    6.7 ^{+   1.4}_{  -1.2}$  & $    4.3 ^{+   0.9}_{  -1.0}$  & $    8.1 ^{+   1.4}_{  -1.2}$  & $    6.3 ^{+   1.7}_{  -1.4}$  & $    4.1 ^{+   1.0}_{  -0.9}$  & $    5.4 ^{+   1.1}_{  -0.9}$  & $    2.7 ^{+   1.0}_{  -0.8}$  & $    3.5 ^{+   0.9}_{  -0.8}$  \\
  6839 & 2006-Nov-29 & $ 7.87 $ & $   105 ^{+   5.1}_{  -6.7}$  & $   81.4 ^{+   7.5}_{  -8.7}$  & $   44.3 ^{+   2.9}_{  -3.8}$  & $   33.4 ^{+   2.8}_{  -2.8}$  & $   28.2 ^{+   3.2}_{  -3.1}$  & $   14.0 ^{+   1.5}_{  -1.4}$  & $   23.5 ^{+   2.3}_{  -2.2}$  & $   21.7 ^{+   2.8}_{  -2.6}$  & $    8.5 ^{+   1.3}_{  -1.2}$  & $   21.4 ^{+   2.5}_{  -2.2}$  & $   18.7 ^{+   2.6}_{  -2.3}$  & $    7.8 ^{+   1.2}_{  -1.1}$  \\
  6840 & 2007-Jan-15 & $ 7.98 $ & $   80.4 ^{+   3.8}_{  -4.0}$  & $   76.6 ^{+   5.8}_{  -6.4}$  & $   26.5 ^{+   2.8}_{  -4.0}$  & $   24.2 ^{+   2.3}_{  -2.1}$  & $   23.3 ^{+   3.0}_{  -3.4}$  & $    8.0 ^{+   1.3}_{  -1.2}$  & $   21.5 ^{+   2.3}_{  -2.2}$  & $   24.1 ^{+   3.0}_{  -2.9}$  & $    7.5 ^{+   1.1}_{  -1.1}$  & $   18.1 ^{+   2.0}_{  -2.0}$  & $   15.0 ^{+   2.2}_{  -1.9}$  & $    8.1 ^{+   1.5}_{  -1.4}$  \\
 11534 & 2010-Jan-01 & $ 28.5 $ & $   79.9 ^{+   2.9}_{  -2.4}$  & $   72.3 ^{+   4.0}_{  -7.2}$  & $   33.5 ^{+   2.9}_{  -1.6}$  & $   21.6 ^{+   1.1}_{  -1.0}$  & $   18.1 ^{+   1.5}_{  -2.1}$  & $    9.2 ^{+   1.3}_{  -0.7}$  & $    8.3 ^{+   0.8}_{  -0.7}$  & $    6.8 ^{+   0.8}_{  -0.8}$  & $    3.6 ^{+   0.5}_{  -0.5}$  & $   48.6 ^{+   1.6}_{  -2.2}$  & $   33.6 ^{+   2.2}_{  -1.9}$  & $   20.5 ^{+   1.1}_{  -1.7}$  \\
 11535 & 2010-Apr-25 & $ 29.4 $ & $   17.3 ^{+   1.1}_{  -1.4}$  & $   10.6 ^{+   1.2}_{  -2.2}$  & $    9.0 ^{+   0.6}_{  -0.6}$  & $    4.4 ^{+   0.5}_{  -0.5}$  & $    2.9 ^{+   0.5}_{  -0.6}$  & $    2.1 ^{+   0.3}_{  -0.3}$  & $    2.3 ^{+   0.4}_{  -0.3}$  & $    1.0 ^{+   0.3}_{  -0.3}$  & $    1.4 ^{+   0.3}_{  -0.2}$  & $    6.7 ^{+   0.7}_{  -0.6}$  & $    3.6 ^{+   0.9}_{  -0.5}$  & $    3.5 ^{+   0.4}_{  -0.4}$  \\
 11536 & 2010-Jun-27 & $ 27.9 $ & $   18.0 ^{+   1.1}_{  -4.1}$  & $    9.4 ^{+   1.1}_{  -1.3}$  & $    9.5 ^{+   0.8}_{  -1.0}$  & $    4.8 ^{+   0.5}_{  -0.5}$  & $    3.2 ^{+   0.5}_{  -0.7}$  & $    2.4 ^{+   0.3}_{  -0.3}$  & $    2.4 ^{+   0.4}_{  -0.3}$  & $    1.3 ^{+   0.4}_{  -0.4}$  & $    1.3 ^{+   0.3}_{  -0.2}$  & $    8.5 ^{+   0.7}_{  -0.7}$  & $    4.3 ^{+   0.7}_{  -0.6}$  & $    4.9 ^{+   0.5}_{  -0.5}$  \\
 11537 & 2010-Aug-08 & $ 29.4 $ & $   10.7 ^{+   1.0}_{  -1.1}$  & $    6.7 ^{+   1.4}_{  -1.0}$  & $    5.4 ^{+   0.5}_{  -0.7}$  & $    2.9 ^{+   0.5}_{  -0.3}$  & $    1.7 ^{+   0.4}_{  -0.5}$  & $    1.5 ^{+   0.3}_{  -0.2}$  & $    1.9 ^{+   0.3}_{  -0.3}$  & $    0.7 ^{+   0.4}_{  -0.2}$  & $    1.2 ^{+   0.3}_{  -0.2}$  & $    4.8 ^{+   0.5}_{  -0.4}$  & $    2.1 ^{+   0.6}_{  -0.4}$  & $    3.0 ^{+   0.4}_{  -0.4}$  \\
 11538 & 2010-Oct-02 & $ 29.4 $ & $   23.4 ^{+   1.4}_{  -1.3}$  & $   15.6 ^{+   1.6}_{  -1.7}$  & $   11.1 ^{+   1.2}_{  -1.2}$  & $    7.2 ^{+   0.6}_{  -0.6}$  & $    4.2 ^{+   0.6}_{  -0.6}$  & $    3.0 ^{+   0.5}_{  -0.3}$  & $    2.5 ^{+   0.4}_{  -0.3}$  & $    1.9 ^{+   0.4}_{  -0.4}$  & $    1.1 ^{+   0.2}_{  -0.2}$  & $   20.1 ^{+   1.1}_{  -1.1}$  & $   13.1 ^{+   1.4}_{  -1.1}$  & $    9.8 ^{+   1.1}_{  -0.8}$  \\
 11539 & 2010-Nov-24 & $ 9.83 $ & $   13.1 ^{+   2.1}_{  -2.2}$  & $    5.7 ^{+   1.4}_{  -1.2}$  & $    7.5 ^{+   1.1}_{  -1.0}$  & $    4.3 ^{+   0.9}_{  -0.9}$  & $    0.8 ^{+   0.5}_{  -0.3}$  & $    3.2 ^{+   0.6}_{  -0.8}$  & $    1.8 ^{+   0.6}_{  -0.5}$  & $    1.3 ^{+   0.7}_{  -0.5}$  & $    0.9 ^{+   0.4}_{  -0.3}$  & $    5.2 ^{+   1.2}_{  -0.9}$  & $    3.7 ^{+   1.3}_{  -0.9}$  & $    2.2 ^{+   0.6}_{  -0.5}$  \\
 13191 & 2010-Nov-27 & $ 9.83 $ & $   11.9 ^{+   1.6}_{  -2.0}$  & $    5.0 ^{+   1.4}_{  -1.0}$  & $    6.8 ^{+   1.0}_{  -1.0}$  & $    2.8 ^{+   0.8}_{  -0.6}$  & $    1.7 ^{+   0.7}_{  -0.6}$  & $    1.5 ^{+   0.5}_{  -0.4}$  & $    1.4 ^{+   0.5}_{  -0.4}$  & $    1.1 ^{+   0.7}_{  -0.5}$  & $    0.6 ^{+   0.3}_{  -0.2}$  & $    3.9 ^{+   1.2}_{  -0.8}$  & $    2.0 ^{+   0.9}_{  -0.7}$  & $    2.3 ^{+   0.6}_{  -0.5}$  \\
 13195 & 2010-Nov-26 & $ 9.83 $ & $   12.8 ^{+   1.6}_{  -4.0}$  & $    6.2 ^{+   2.0}_{  -1.1}$  & $    6.9 ^{+   0.9}_{  -1.2}$  & $    3.4 ^{+   0.8}_{  -1.3}$  & $    1.2 ^{+   0.6}_{  -0.4}$  & $    2.1 ^{+   0.5}_{  -0.5}$  & $    1.5 ^{+   0.6}_{  -0.5}$  & $    1.0 ^{+   0.6}_{  -0.4}$  & $    0.8 ^{+   0.4}_{  -0.3}$  & $    4.0 ^{+   1.3}_{  -0.8}$  & $    2.1 ^{+   0.9}_{  -0.7}$  & $    2.2 ^{+   0.6}_{  -0.5}$  \\
 13960 & 2012-Jan-10 & $ 29.4 $ & $   15.2 ^{+   0.9}_{  -1.1}$  & $    7.7 ^{+   1.1}_{  -1.4}$  & $    8.3 ^{+   0.6}_{  -1.0}$  & $    5.5 ^{+   0.6}_{  -0.5}$  & $    2.4 ^{+   0.5}_{  -0.4}$  & $    3.0 ^{+   0.4}_{  -0.6}$  & $    2.5 ^{+   0.4}_{  -0.4}$  & $    1.1 ^{+   0.4}_{  -0.3}$  & $    1.5 ^{+   0.3}_{  -0.3}$  & $    5.2 ^{+   0.6}_{  -0.6}$  & $    3.4 ^{+   0.8}_{  -0.7}$  & $    2.5 ^{+   0.7}_{  -0.4}$  \\
 13961 & 2012-Aug-03 & $ 29.2 $ & $   39.8 ^{+   2.2}_{  -2.9}$  & $   30.1 ^{+   3.3}_{  -3.3}$  & $   17.2 ^{+   1.1}_{  -1.4}$  & $   11.6 ^{+   0.8}_{  -0.9}$  & $    8.1 ^{+   1.2}_{  -0.8}$  & $    5.1 ^{+   0.5}_{  -0.5}$  & $    9.5 ^{+   0.7}_{  -0.7}$  & $    8.8 ^{+   1.0}_{  -1.0}$  & $    3.9 ^{+   0.4}_{  -0.4}$  & $   10.6 ^{+   0.9}_{  -0.8}$  & $    8.8 ^{+   1.0}_{  -0.9}$  & $    4.4 ^{+   0.5}_{  -0.4}$  \\
 14513 & 2012-Dec-26 & $ 28.6 $ & $   34.2 ^{+   1.7}_{  -2.1}$  & $   24.7 ^{+   1.9}_{  -2.4}$  & $   14.6 ^{+   1.5}_{  -1.4}$  & $   11.1 ^{+   0.8}_{  -0.7}$  & $    7.3 ^{+   0.9}_{  -0.9}$  & $    6.2 ^{+   0.5}_{  -0.8}$  & $   17.0 ^{+   1.2}_{  -1.2}$  & $   12.7 ^{+   1.3}_{  -1.2}$  & $    8.3 ^{+   0.7}_{  -1.6}$  & $   11.6 ^{+   1.0}_{  -0.9}$  & $    6.7 ^{+   0.8}_{  -0.7}$  & $    6.6 ^{+   0.7}_{  -0.7}$  \\
 14514 & 2013-Jan-06 & $ 29.4 $ & $   32.4 ^{+   1.4}_{  -1.8}$  & $   22.3 ^{+   2.3}_{  -2.5}$  & $   14.9 ^{+   0.9}_{  -0.9}$  & $    9.9 ^{+   0.8}_{  -0.8}$  & $    6.1 ^{+   0.8}_{  -0.7}$  & $    5.4 ^{+   0.5}_{  -0.5}$  & $   15.9 ^{+   1.2}_{  -1.1}$  & $   11.5 ^{+   1.0}_{  -1.0}$  & $    8.5 ^{+   0.7}_{  -0.6}$  & $   12.1 ^{+   0.8}_{  -0.8}$  & $    8.2 ^{+   1.0}_{  -0.9}$  & $    6.1 ^{+   0.5}_{  -0.5}$  \\
 14515 & 2013-Aug-31 & $ 9.73 $ & $   21.1 ^{+   3.2}_{  -3.4}$  & $   10.7 ^{+   2.0}_{  -1.6}$  & $   12.3 ^{+   1.6}_{  -2.2}$  & $    6.8 ^{+   2.4}_{  -0.9}$  & $    4.3 ^{+   1.1}_{  -0.9}$  & $    3.8 ^{+   0.9}_{  -0.8}$  & $    7.2 ^{+   1.6}_{  -1.1}$  & $    5.1 ^{+   1.6}_{  -1.1}$  & $    3.8 ^{+   0.8}_{  -0.7}$  & $    4.5 ^{+   1.0}_{  -1.0}$  & $    4.0 ^{+   1.2}_{  -0.9}$  & $    1.6 ^{+   0.6}_{  -0.5}$  \\
 14516 & 2013-Oct-01 & $ 29.4 $ & $   13.9 ^{+   0.9}_{  -2.3}$  & $    7.5 ^{+   1.1}_{  -1.3}$  & $    7.5 ^{+   0.6}_{  -0.7}$  & $    5.2 ^{+   0.6}_{  -0.5}$  & $    2.3 ^{+   0.5}_{  -0.4}$  & $    2.9 ^{+   0.4}_{  -0.4}$  & $    4.2 ^{+   0.5}_{  -0.5}$  & $    2.3 ^{+   0.5}_{  -0.5}$  & $    2.1 ^{+   0.3}_{  -0.3}$  & $    3.0 ^{+   0.4}_{  -0.4}$  & $    1.6 ^{+   0.4}_{  -0.4}$  & $    1.7 ^{+   0.3}_{  -0.3}$  \\
 14517 & 2014-May-15 & $ 29.4 $ & $   49.5 ^{+   2.3}_{  -9.2}$  & $   34.0 ^{+   2.0}_{  -2.2}$  & $   24.2 ^{+   1.4}_{  -1.6}$  & $   17.5 ^{+   1.3}_{  -1.4}$  & $   13.4 ^{+   1.2}_{  -1.1}$  & $    8.4 ^{+   0.7}_{  -0.7}$  & $   16.0 ^{+   0.9}_{  -0.9}$  & $   10.4 ^{+   1.0}_{  -1.0}$  & $    8.1 ^{+   0.6}_{  -0.6}$  & $    9.1 ^{+   0.8}_{  -0.7}$  & $    5.5 ^{+   0.7}_{  -0.6}$  & $    4.2 ^{+   0.4}_{  -0.4}$  \\
 14518 & 2014-Jun-08 & $ 29.3 $ & $   28.8 ^{+   7.5}_{  -3.0}$  & $   22.3 ^{+   1.8}_{  -2.6}$  & $   14.6 ^{+   2.7}_{  -1.0}$  & $   11.1 ^{+   0.9}_{  -2.6}$  & $    6.5 ^{+   1.1}_{  -0.7}$  & $    5.0 ^{+   0.6}_{  -0.5}$  & $   10.5 ^{+   0.8}_{  -3.3}$  & $    6.5 ^{+   0.9}_{  -0.8}$  & $    5.0 ^{+   0.6}_{  -0.9}$  & $    6.0 ^{+   0.7}_{  -2.2}$  & $    3.4 ^{+   0.6}_{  -0.5}$  & $    3.1 ^{+   0.4}_{  -0.5}$  \\
 16316 & 2013-Aug-26 & $ 9.83 $ & $   15.6 ^{+   2.9}_{  -1.4}$  & $   10.9 ^{+   2.3}_{  -2.1}$  & $    8.1 ^{+   1.2}_{  -2.6}$  & $    5.5 ^{+   1.1}_{  -0.8}$  & $    3.2 ^{+   1.4}_{  -0.7}$  & $    3.1 ^{+   1.4}_{  -0.6}$  & $    6.3 ^{+   1.3}_{  -1.2}$  & $    2.5 ^{+   1.1}_{  -0.7}$  & $    3.5 ^{+   2.0}_{  -0.6}$  & $    5.1 ^{+   1.0}_{  -1.3}$  & $    3.2 ^{+   1.1}_{  -0.8}$  & $    1.9 ^{+   1.3}_{  -0.4}$  \\
 16317 & 2013-Aug-29 & $ 9.83 $ & $   14.6 ^{+   1.9}_{  -2.0}$  & $    9.0 ^{+   1.9}_{  -1.5}$  & $    7.4 ^{+   1.0}_{  -1.1}$  & $    5.8 ^{+   1.1}_{  -1.0}$  & $    3.9 ^{+   1.2}_{  -0.9}$  & $    2.7 ^{+   0.6}_{  -0.6}$  & $    6.6 ^{+   1.0}_{  -1.0}$  & $    2.8 ^{+   1.0}_{  -0.9}$  & $    3.9 ^{+   0.7}_{  -0.7}$  & $    3.6 ^{+   0.8}_{  -0.7}$  & $    2.7 ^{+   0.9}_{  -0.8}$  & $    1.5 ^{+   0.5}_{  -0.4}$  \\
\enddata

\tablecomments{Count rates are in units of $10^{-3}\,{\rm s}^{-1}$. \textit{Exp} reports the values stored in the header keyword EXPOSURE in units of $10^{3}\,{\rm s}$}

\end{deluxetable}



\begin{deluxetable}{lllllll}
\tablewidth{0pt}
\tabletypesize{\scriptsize}
\tablecolumns{4}


\centering

\tablecaption{Lens model properties at the images \label{tab:kappasgammas}}


\tablehead{
\colhead{Object} &
\colhead{Image} &
\colhead{$R/R_{ef}$} &
\colhead{$\kappa_{*} / \kappa$ } &
\colhead{$\kappa$ } &
\colhead{$\gamma$ } &
\colhead{Macrolens model}
}

\startdata

QJ 0158$-$4325  &  A  &  1.23  &  0.39  &  0.348  &  0.428  &  SIE+g  \\
QJ 0158$-$4325  &  B  &  0.62  &  0.63  &  0.693  &  0.774  &  SIE+g  \\
HE 0435$-$1223  &  A  &  1.71  &  0.27  &  0.445  &  0.383  &  SIE+g  \\
HE 0435$-$1223  &  B  &  1.54  &  0.31  &  0.539  &  0.602  &  SIE+g  \\
HE 0435$-$1223  &  C  &  1.71  &  0.27  &  0.444  &  0.396  &  SIE+g  \\
HE 0435$-$1223  &  D  &  1.40  &  0.34  &  0.587  &  0.648  &  SIE+g  \\
SDSS 0924+0219  &  A  &  2.93  &  0.12  &  0.472  &  0.456  &  SIE+g  \\
SDSS 0924+0219  &  B  &  3.26  &  0.10  &  0.443  &  0.383  &  SIE+g  \\
SDSS 0924+0219  &  C  &  2.69  &  0.14  &  0.570  &  0.591  &  SIE+g  \\
SDSS 0924+0219  &  D  &  2.79  &  0.13  &  0.506  &  0.568  &  SIE+g  \\
SDSS 1004+4112  &  A  &  $-$   &  0.03  &  0.763  &  0.300  &  parametric    \\
SDSS 1004+4112  &  B  &  $-$   &  0.03  &  0.696  &  0.204  &  parametric    \\
SDSS 1004+4112  &  C  &  $-$   &  0.03  &  0.635  &  0.218  &  parametric    \\
SDSS 1004+4112  &  D  &  $-$   &  0.03  &  0.943  &  0.421  &  parametric    \\
HE 1104$-$1805  &  A  &  1.70  &  0.27  &  0.610  &  0.512  &  SIE+g  \\
HE 1104$-$1805  &  B  &  3.29  &  0.09  &  0.321  &  0.217  &  SIE+g  \\
Q 2237+0305     &  A  &  0.24  &  0.79  &  0.39   &  0.40   &  SIE+g  \\
Q 2237+0305     &  B  &  0.25  &  0.79  &  0.38   &  0.39   &  SIE+g  \\
Q 2237+0305     &  C  &  0.20  &  0.81  &  0.74   &  0.73   &  SIE+g  \\
Q 2237+0305     &  D  &  0.23  &  0.80  &  0.64   &  0.62   &  SIE+g  \\
\enddata

\tablecomments{For each image we give the distance $R/R_{ef}$ of the image from the lens center in units of the effective radius of the lens from \cite{Oguri2014}, the expected fraction of the surface density in stars $\kappa_{*}/\kappa$, the surface density in stars, the surface density $\kappa$ in units of the lens critical density and the total shear $\gamma$. For QJ 0158$-$4325 and HE 1104$-$1805 we used our own model. We used the models of \cite{Schechter2014} for HE 0435$-$1223 and SDSS 0924+0219, \cite{Kochanek2004} for Q 2237+0305, and \cite{Oguri2014} for SDSS 1004+4112}

\end{deluxetable}


\begin{deluxetable}{llllllllllll}
\tablewidth{0pt}
\tabletypesize{\scriptsize}
\tablecolumns{4}
 

\centering
  
\tablecaption{Time-averaged microlensing magnifications (magnitude)\label{tab:microlensingtable}}
  
  
\tablehead{
\colhead{Object} &
\colhead{Pair} &
\colhead{Epochs} &
\colhead{$\Delta m^{full}$} &
\colhead{$\Delta m^{soft}$} &
\colhead{$\Delta m^{hard}$}
}
  
  \startdata
  
  QJ 0158$-$4325 & (B-A) & 12 & $ +0.37 \pm  0.17 $ & $ +0.25 \pm  0.19 $ & $ +0.45 \pm  0.23 $  \\
  HE 0435$-$1223 & (B-A) & 10 & $ +0.55 \pm  0.13 $ & $ +0.58 \pm  0.17 $ & $ +0.58 \pm  0.18 $  \\
  HE 0435$-$1223 & (C-A) & 10 & $ +0.26 \pm  0.12 $ & $ +0.36 \pm  0.17 $ & $ +0.21 \pm  0.16 $  \\
  HE 0435$-$1223 & (D-A) & 10 & $ -0.04 \pm  0.13 $ & $ +0.01 \pm  0.17 $ & $ -0.07 \pm  0.17 $  \\
  HE 0435$-$1223 & (C-B) & 10 & $ -0.29 \pm  0.14 $ & $ -0.22 \pm  0.18 $ & $ -0.37 \pm  0.17 $  \\
  HE 0435$-$1223 & (D-B) & 10 & $ -0.59 \pm  0.15 $ & $ -0.56 \pm  0.18 $ & $ -0.64 \pm  0.18 $  \\
  HE 0435$-$1223 & (D-C) & 10 & $ -0.30 \pm  0.13 $ & $ -0.34 \pm  0.18 $ & $ -0.27 \pm  0.16 $  \\
  SDSS 0924+0219 & (B-A) &  6 & $ +0.67 \pm  0.20 $ & $ +0.69 \pm  0.23 $ & $ +0.74 \pm  0.34 $  \\
  SDSS 0924+0219 & (C-A) &  6 & $ +1.76 \pm  0.27 $ & $ +1.81 \pm  0.32 $ & $ +2.08 \pm  0.58 $  \\
  SDSS 0924+0219 & (D-A) &  6 & $ +2.26 \pm  0.28 $ & $ +2.11 \pm  0.30 $ & $ +2.59 \pm  0.58 $  \\
  SDSS 0924+0219 & (C-B) &  6 & $ +1.09 \pm  0.30 $ & $ +1.12 \pm  0.35 $ & $ +1.34 \pm  0.64 $  \\
  SDSS 0924+0219 & (D-B) &  6 & $ +1.59 \pm  0.31 $ & $ +1.42 \pm  0.33 $ & $ +1.86 \pm  0.64 $  \\
  SDSS 0924+0219 & (D-C) &  6 & $ +0.50 \pm  0.36 $ & $ +0.30 \pm  0.41 $ & $ +0.52 \pm  0.79 $  \\
  SDSS 1004+4112 & (B-A) & 11 & $ -0.99 \pm  0.09 $ & $ -1.06 \pm  0.13 $ & $ -0.89 \pm  0.13 $  \\
  SDSS 1004+4112 & (C-A) & 11 & $ -1.51 \pm  0.09 $ & $ -1.56 \pm  0.13 $ & $ -1.45 \pm  0.13 $  \\
  SDSS 1004+4112 & (D-A) & 11 & $ -1.70 \pm  0.10 $ & $ -1.73 \pm  0.15 $ & $ -1.65 \pm  0.14 $  \\
  SDSS 1004+4112 & (C-B) & 11 & $ -0.52 \pm  0.08 $ & $ -0.49 \pm  0.11 $ & $ -0.56 \pm  0.12 $  \\
  SDSS 1004+4112 & (D-B) & 11 & $ -0.71 \pm  0.09 $ & $ -0.66 \pm  0.13 $ & $ -0.76 \pm  0.13 $  \\
  SDSS 1004+4112 & (D-C) & 11 & $ -0.19 \pm  0.09 $ & $ -0.17 \pm  0.13 $ & $ -0.20 \pm  0.13 $  \\
  HE 1104$-$1805 & (B-A) & 15 & $ -1.42 \pm  0.16 $ & $ -1.61 \pm  0.28 $ & $ -1.32 \pm  0.20 $  \\
  Q 2237+0305    & (B-A) & 30 & $ +1.13 \pm  0.14 $ & $ +1.22 \pm  0.20 $ & $ +1.11 \pm  0.17 $  \\
  Q 2237+0305    & (C-A) & 30 & $ +0.66 \pm  0.15 $ & $ +0.66 \pm  0.22 $ & $ +0.65 \pm  0.19 $  \\
  Q 2237+0305    & (D-A) & 30 & $ +1.16 \pm  0.15 $ & $ +1.21 \pm  0.22 $ & $ +1.15 \pm  0.18 $  \\
  Q 2237+0305    & (C-B) & 30 & $ -0.47 \pm  0.16 $ & $ -0.56 \pm  0.25 $ & $ -0.46 \pm  0.21 $  \\
  Q 2237+0305    & (D-B) & 30 & $ +0.03 \pm  0.16 $ & $ -0.01 \pm  0.25 $ & $ +0.04 \pm  0.20 $  \\
  Q 2237+0305    & (D-C) & 30 & $ +0.50 \pm  0.17 $ & $ +0.55 \pm  0.27 $ & $ +0.50 \pm  0.22 $  \\
 
  \enddata
\end{deluxetable}


\begin{deluxetable}{llll}
\tablewidth{0pt}
\tabletypesize{\scriptsize}
\tablecolumns{4}
 

\centering
  
\tablecaption{Upper limit estimates in units of gravitational radii. \label{tab:softandhardresults_gr}}
  
  
\tablehead{
\colhead{$ $} &
\colhead{$p=68\%$} &
\colhead{$p=95\%$} &
\colhead{$p=99\%$}
}
\startdata
 Soft X-rays  & $ 17.8 $ & $ 39.5 $ & $ 73.0 $  \\
 Hard X-rays  & $ 18.9 $ & $ 42.3 $ & $ 97.1 $  \\
 Full         & $ 16.7 $ & $ 36.1 $ & $ 62.1 $  \\
 \textit{Full, single-epoch PDF} & $ 28.9 $ & $ 58.8 $ & $ 99.3 $  \\
 
  \enddata
\tablecomments{The last line contains the estimates when no correction is done for the length of the observation campaings, hence comparing averaged curves against single-epoch probability distributions.}
  \end{deluxetable}

\end{document}